\documentclass[twocolumn]{raa}
 
\usepackage{graphicx,times}             
\usepackage{natbib}
\usepackage{amssymb,amsmath}
\usepackage{mathrsfs}
\bibpunct{(}{)}{;}{a}{}{,}

\usepackage[pagebackref=true]{hyperref}

\def\aj{AJ}

\def\apjs{ApJS}

\def\mnras{MNRAS}
\def\aap{A\&A}

\def\nat{Nature}      
\def\apjs{ApJS}
\def\apjl{ApJ Letters}

\begin{document}

\title{Do minor interactions trigger star formation in galaxy pairs?}

   \volnopage{Vol.0 (20xx) No.0, 000--000}     
   \setcounter{page}{1}           

   \author{Apashanka Das
      \inst{1}
   \and Biswajit Pandey
      \inst{2}
   \and Suman Sarkar
      \inst{3,4}
   }
 \institute{Department of Physics, Visva-Bharati University, Santiniketan, 
              Birbhum, 731235, India
             {\it a.das.cosmo@gmail.com}\\
        \and
	     {Department of Physics, Visva-Bharati University, Santiniketan, 
	     Birbhum, 731235, India {\it biswap@visva-bharati.ac.in}}\\
        \and
            {Department of Physics, Indian Institute of Science Education and
              Research Tirupati, Tirupati - 517507, India
        \and
            {Department of Physics, Indian Institute of Technology Kharagpur, Kharagpur, 721302, India}
              {\it suman2reach@gmail.com}}
   }
\vs\no
   {\small Received 20xx month day; accepted 20xx month day}

\abstract{We analyze the galaxy pairs in a set of volume limited
  samples from the SDSS to study the effects of minor interactions on
  the star formation rate (SFR) and colour of galaxies. We carefully
  design control samples of the isolated galaxies by matching the
  stellar mass and redshift of the minor pairs. The SFR distributions
  and colour distributions in the minor pairs differ from their
  controls at $>99\%$ significance level. We also simultaneously match
  the control galaxies in stellar mass, redshift and local density to
  assess the role of the environment. The null hypothesis can be
  rejected at $>99\%$ confidence level even after matching the
  environment. Our analysis shows a quenching in the minor pairs where
  the degree of quenching decreases with the increasing pair
  separation and plateaus beyond 50 kpc. We also prepare a sample of
  minor pairs with $H_{\alpha}$ line information. We calculate the SFR
  of these galaxies using the $H_{\alpha}$ line and repeat our
  analysis. We observe a quenching in the $H_{\alpha}$ sample too. We
  find that the majority of the minor pairs are quiescent systems that
  could be quenched due to minor interactions. Combining data from the
  Galaxy Zoo and Galaxy Zoo2, we find that only $\sim 1\%$ galaxies
  have a dominant bulge, $4\%-7\%$ galaxies host a bar, and $5\%-10\%$
  galaxies show the AGN activity in minor pairs. This indicates that
  the presence of bulge, bar and AGN activity plays an insignificant
  role in quenching the galaxies in minor pairs. The more massive
  companion satisfies the criteria for mass quenching in most of the
  minor pairs. We propose that the stripping and starvation likely
  caused the quenching in the less massive companion at a later stage
  of evolution. \keywords{methods: statistical -- data analysis --
    galaxies: formation -- evolution -- cosmology: large scale
    structure of the Universe.}}

\authorrunning{A. Das, B. Pandey \& S. Sarkar}
\titlerunning{Do minor interactions trigger star formation in galaxy pairs?}
 
\maketitle

\section{Introduction}
The formation and the evolution of the galaxies in the Universe
remains one of the most challenging problems in modern cosmology. The
$\Lambda$CDM model is quite successful in explaining most of the
cosmological observations on large-scales. However, our understanding
of the details of the galaxy formation and evolution within the
framework of the $\Lambda$CDM model is still incomplete. The first
bound objects form in the Universe when the primordial density
fluctuations in the dark matter density field collapse into dark
matter halos. The galaxies are believed to have formed by the
accretion of neutral hydrogen gas onto the dark matter halos and the
subsequent cooling and condensation of the gas at the centre of these
halos \citep{reesostriker77, silk77, white78, fall80}.

The galaxies are not island universes that evolve in isolation. They
are an integral part of an extensive and complex network, namely the
cosmic web \citep{bond96}. The initial conditions at the location of
formation, the assembly history and the interactions with the
environment may play crucial roles in the formation and evolution of a
galaxy. In the hierarchical scenario, the galaxy interactions and
mergers provide an efficient mechanism for the buildup of massive
galaxies. Such processes can modify the mass distribution, morphology
and the star formation activity in the galaxies. Using simulations of
tidal interactions, \citet{toomre72} first show that the spiral and
irregular galaxies could transform into ellipticals and S0
galaxies. Subsequent studies with more sophisticated simulations
\citep{barnes96, mihos96, tissera02, cox06, dimatteo07, montuori10,
  rupke10, torrey12, renaud14} reveal that the tidal torques generated
during the encounter can trigger starburst in the interacting
galaxies. The efficiency of the tidally triggered star formation is
known to depend on several factors such as the amount of available
gas, depth of the potential well, morphology, orbital parameters and
the internal dynamical properties of the galaxies \citep{barnes96,
  tissera00, perez06}.

The first observational evidence of enhanced star formation in
interacting galaxies came from a seminal study of optical colours in
the morphologically disturbed galaxies by
\citet{larson78}. Subsequently, many other studies on interacting
galaxy pairs from the modern spectroscopic redshift surveys confirm
the SFR enhancement at smaller pair separation \citep{barton00,
  lambas03, alonso04, nikolic04, alonso06, woods06, woods07, barton07,
  ellison08, heiderman09, knapen09, robaina09, ellison10, woods10,
  patton11, pan18, das23}. The level of enhancement reported in most
of these studies is within a factor of two compared to the isolated
galaxies. The enhancement is known to depend on multiple factors such
as the separation, luminosity or mass ratio and the type of galaxies
involved in the interaction. The changes in the star formation is very
often studied as a function of the projected separation because it is
believed as an indicator of the merger phase of a galaxy pair. The
pairs at smaller separation are most likely undergoing a close
passage.  On the other hand, the pairs at larger separation may be
approaching each other or receding away after their first pericentric
passage. Nevertheless, these are difficult to confirm as the projected
separation corresponds to a snapshot view of the interaction that does
not provide any direct information about the time scale.

 The equilibrium model \citep{dekel09, bouche10,
    dave11, dave12, lilly13} emerged as a successful model of galaxy
  evolution over the last decade. The model suggests that the galaxies
  maintain an equilibrium between inflow, outflow and star
  formation. The galaxies are perturbed off the equilibrium relations
  by interactions and mergers. These galaxies tend to be driven back
  towards equilibrium. The deviations in SFR from the equilibrium
  relation are correlated with the available gas fraction. So the star
  formation rate of a galaxy is primarily decided by the available gas
  mass, which itself is modulated by inflows and outflows of
  gas. Besides, both the fraction of molecular gas in the cold ISM
  reservoir and the rate of conversion of molecular gas to stars are
  also important \citep{saintonge22}. \citet{saintonge12} show that
  the global SFR is driven by both the molecular gas density and the
  gas depletion time. The molecular gas properties may be also
  affected by galaxy interactions. \citet{pan18} find that
  interactions modify the molecular gas properties in galaxy pairs
  where the magnitude of the effect is sensitive to the pair
  configuration. \citet{elli18} study the atomic Hydrogen gas fraction
  in post-merger galaxies and find that the enhanced atomic gas
  fractions in post-mergers are not a consequence of the merger
  induced star bursts or outflows but arises due to the enhanced
  turbulence that decreases the star formation
  efficiency. \citet{thorp22} study merger induced starbursts using
  ALMaQUEST survey and find that the star formation in some mergers
  are driven by the abundance of molecular gas fuel whereas the star
  formation efficiency plays a leading role in
  others. \citet{violino18} study the relation between star formation
  and molecular gas properties in galaxy mergers and find that both
  interactions and internal processes may lead to molecular gas
  enhancement and decreased depletion times. All these studies suggest
  that the available gas mass plays a crucial role for star formation
  in galaxies. However, the galaxies are diverse in their details and
  star formation in interacting galaxies also depend on the properties
  of the interacting pairs. Many of these findings are also supported
  by parsec-scale galaxy-merger simulations \citep{renaud14, moreno21}
  and analysis of galaxy pairs from hydrodynamical simulations
  \citep{patton20}.

The simulations of galaxy interactions suggest that the tidally
triggered star formation is more efficient in galaxy pairs with
similar stellar mass or luminosity. Such interactions are known as the
major interactions. On the other hand, the minor interactions are the
interactions between galaxies with a relatively larger mass or
luminosity ratio. The minor interactions and mergers are expected to
be more frequent in a galaxy's history since the frequency of mergers
of dark matter halos increases with their mass ratio \citep{lacey93,
  fakhouri08}. Studies with simulations \citep{mihos94, mastro05,
  cox08} indicate that a lower level of star formation enhancement may
also occur in minor mergers after several billion years. 

Most of the observational studies of galaxy pairs confirm the tidally
triggered star formation in major interactions. However, the issue of
star formation enhancement in minor interactions are less clear. In
the hierarchical galaxy formation model \citep{somerville99,
  kauffmann1, kauffmann2, diaferio99}, most interactions and mergers
occur between unequal-mass systems due to the greater abundance of low
mass and low luminosity galaxies. The minor interactions may thus play
a crucial role in galaxy evolution. Any observational study of minor
interaction and merger is challenging due to several reasons. The
number of minor pairs identified from magnitude limited surveys are
far less than the number of major pairs as the galaxies have similar
magnitudes in such surveys. It is also difficult to identify the
low-luminosity companions around the more luminous members due to the
contaminations from the background galaxies. Despite these
limitations, the effects of minor interactions on star formation have
drawn considerable interest. \citet{lambas03} study the star formation
enhancement in paired galaxies using 2dFGRS and find a dependence on
the relative luminosity of the pairs. \citet{nikolic04} use SDSS to
analyze the star formation in paired galaxies and find no dependence
on the luminosity of the companion galaxy. \citet{woods06} analyze
data from CfA2 survey and a follow-up search to find that the star
formation enhancement in pairs decreases with increasing stellar mass
ratio. \citet{woods07} show that the specific SFR of the less massive
member in a minor pair is enhanced, whereas the more massive member
remains unaffected. \citet{ellison08} analyze the SDSS data and find
tentative evidence for higher SFR for the less massive companions in
minor pairs at a low significance level. \citet{li08} also reach a
similar conclusion using the SDSS data. Our current understanding of
the impact of minor interactions and mergers are far from
complete. The observational studies do not provide a conclusive
evidence of enhanced star formation in minor pairs in the present
Universe.

The Sloan Digital Sky Survey (SDSS) \citep{strauss02} is the largest
photometric and spectroscopic redshift survey available at present.
The availability of precise spectroscopic information for a large
number of galaxies in the SDSS provides an excellent opportunity for
the statistical study of minor interactions and their effects on the
star formation and colour. We intend to study the SFR and colour of
the minor galaxy pairs in the present Universe using the SDSS.

The galaxy colour is strongly correlated with the star formation due
to the observed bimodality \citep{strateva01, baldry04, pandey20}. The
galaxies in the blue cloud are gas rich and they have higher star
formation rates. Contrarily, the red sequence hosts the gas poor
galaxies with very low to no star formation. The tidal interactions
and mergers between galaxies may trigger starbursts or quenching which
consequently alter their colours. Such colour changes usually happen
on a time scale longer than the starburst or quenching. The effect of
tidal interactions on the galaxy pairs can be captured more
convincingly if we employ both star formation rate and colour for such
studies. We plan to study the star formation rate and the dust
corrected $(u-r)$ colour of the minor galaxy pairs in a set of volume
limited samples from the SDSS and compare these with the respective
control samples of the isolated galaxies.

The outline of the paper is as follows. We describe the data and the
method of analysis in Section 2, discuss the results in Section 3 and
present our conclusions in Section 4.\\

Through out the paper we use the $\Lambda$CDM cosmological model with
$\Omega_{m0}=0.315$, $\Omega_{\Lambda0}=0.685$ and $h=0.674$
\citep{planck18} for conversion of redshift to comoving distance.\\

\section{DATA AND METHOD OF ANALYSIS}   

\subsection{SDSS DR16}

The Sloan Digital Sky Survey (SDSS) is a multi-band imaging and
spectroscopic redshift survey with a 2.5 m telescope at Apache Point
Observatory in New Mexico. The technical details of the SDSS telescope
is described in \citep{gunn06} and a description of the SDSS
photometric camera can be found in \citep{gunn98}. The selection
algorithm for the SDSS Main galaxy sample is discussed in
\citep{strauss02} and a technical summary of the survey is provided in
\citep{york00}.

We use the sixteenth data release (DR16) of the SDSS to identify the
galaxy pairs in the nearby universe. We use the structured query
language (SQL) to download the spectroscopic and photometric
information of galaxies in DR16 \citep{ahumada20} from the SDSS
CasJobs \footnote{https://skyserver.sdss.org/casjobs/}. We select a
contiguous region of the sky that spans $135^{\circ} \leq \alpha \leq
225^{\circ}$ and $0^{\circ} \leq \delta \leq 60^{\circ}$ in equatorial
co-ordinates. We consider all the galaxies with $r$ band Petrosian
magnitudes $m_r \leq 17.77 $ and construct three volume limited
samples by restricting the $r$ band absolute magnitude to $M_r \leq
-19$, $M_r \leq -20$ and $M_r \leq -21$. The details of these samples
are provided in the \autoref{tab1}. The three magnitude bins are not
independent of each other. We also tried to construct pair samples
using independent magnitude bins. However this drastically reduces the
number of available minor pairs. The primary motivation behind the
choice of these magnitude bins is to investigate if there are any
luminosity dependence of the outcomes of minor interactions.

We obtain the stellar mass and the star formation rate (SFR) of the
galaxies from the {\it{StellarMassFSPSGranWideNoDust}} table. These
are calculated using the Flexible Stellar Population Synthesis (FSPS)
model \citep{conroy09}. We retrieve the information of internal
reddening E(B-V) of the galaxies from the {\it{emmissionLinesPort}}
table which is based on publicly available Gas AND Absorption Line
Fitting (GANDALF) \citep{gandalf} and penalised PIXEL Fitting (pPXF)
\citep{ppxf}. We correct for the dust attenuation in the source galaxy
by using its internal reddening E(B-V). While downloading the above
information, we consider only those galaxies which have their
$scienceprimary$ flag set to 1. This ensures that only the galaxies
with high quality spectra are used in our analysis. Further, we obtain
morphological information of the galaxies in three volume limited
samples by cross-matching their $SpecObjID$ with the galaxies in
Galaxy Zoo \citep{lintott08}. We identify elliptical and spiral
galaxies respectively as those which have their elliptical and spiral
debiased vote fraction $>$ 0.8. We also obtain the information about
the presence of a dominant bulge and a bar in the galaxies in our
sample by cross-matching their $SpecObjID$ with the galaxies in Galaxy
Zoo 2 \citep{willett}. We also identify the paired galaxies with an
AGN by cross-matching their $SpecObjID$ with the galaxies present in
the MPA-JHU spectroscopic catalogue \citep{brinchmann04,
  kauffmann03b}.

\begin{figure*}
\resizebox{15cm}{6cm}{\rotatebox{0}{\includegraphics{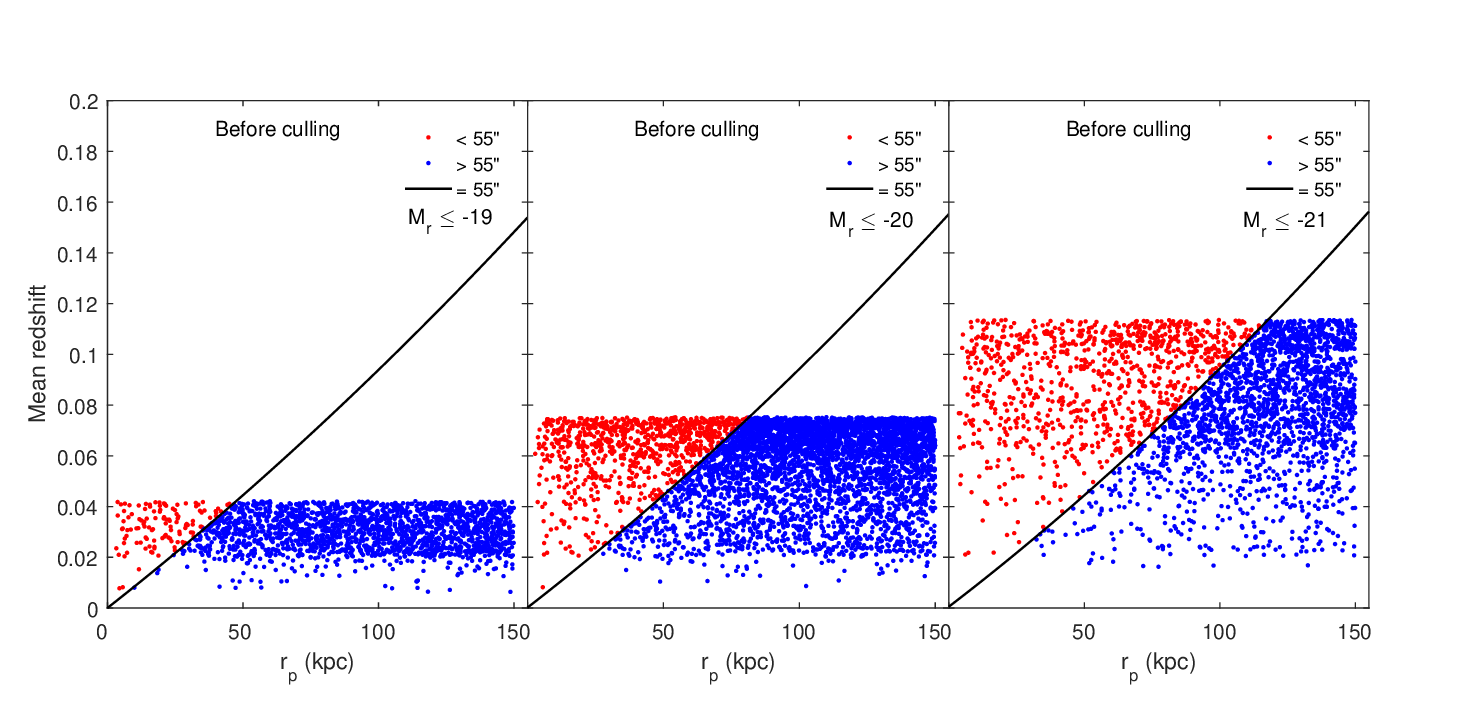}}}
\caption*{}
\end{figure*}

\begin{figure*}
\resizebox{15cm}{6cm}{\rotatebox{0}{\includegraphics{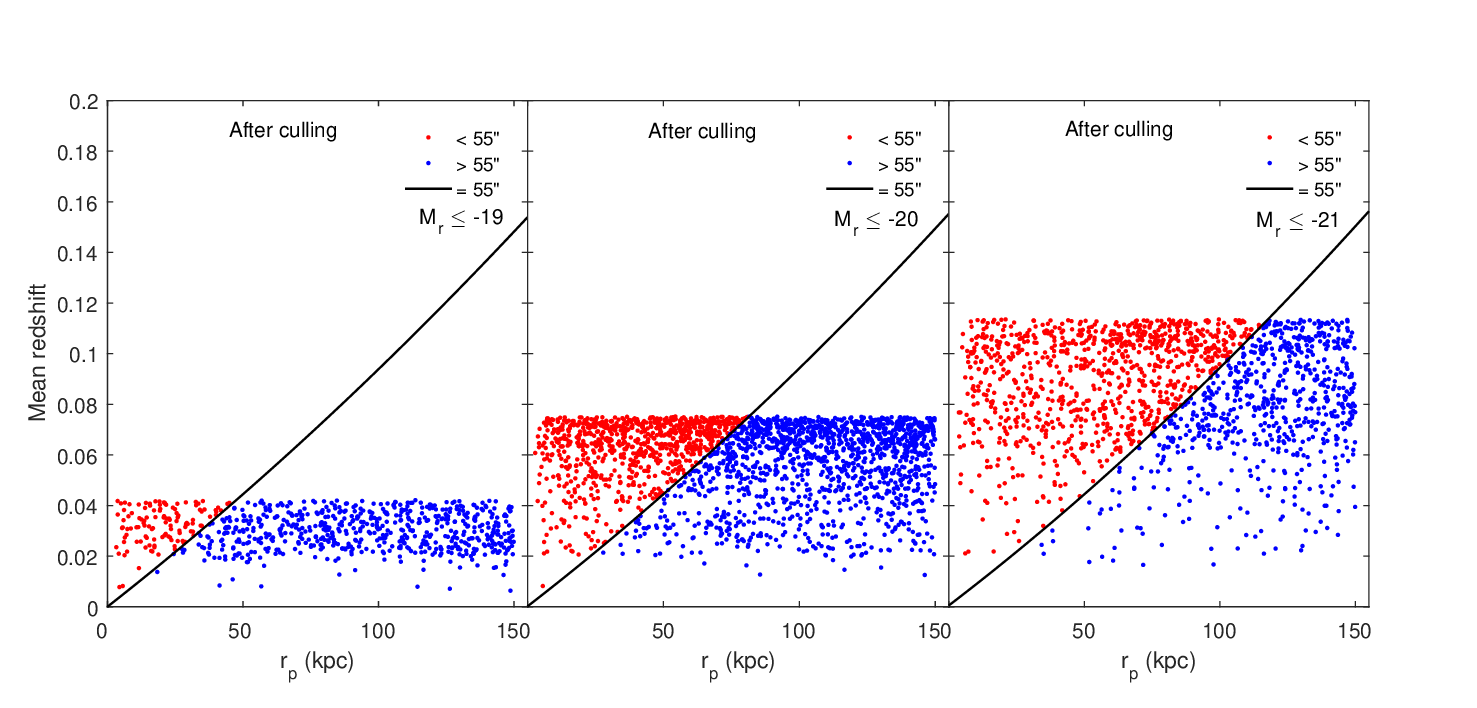}}}

\caption{The top panels show the mean redshift versus the projected
  separation of pairs before culling in the three volume limited
  samples. Pairs having angular separation less than
  $55^{\prime\prime}$ and greater than $55^{\prime\prime}$ are shown
  using red and blue dots respectively. The black solid line
  represents the theoretical curve for angular separation equal to
  $55^{\prime\prime}$. The bottom panels show the same after culling.}

\label{fig:cullpair}
\end{figure*}

\begin{figure*}
\resizebox{15cm}{21cm}{\rotatebox{0}{\includegraphics{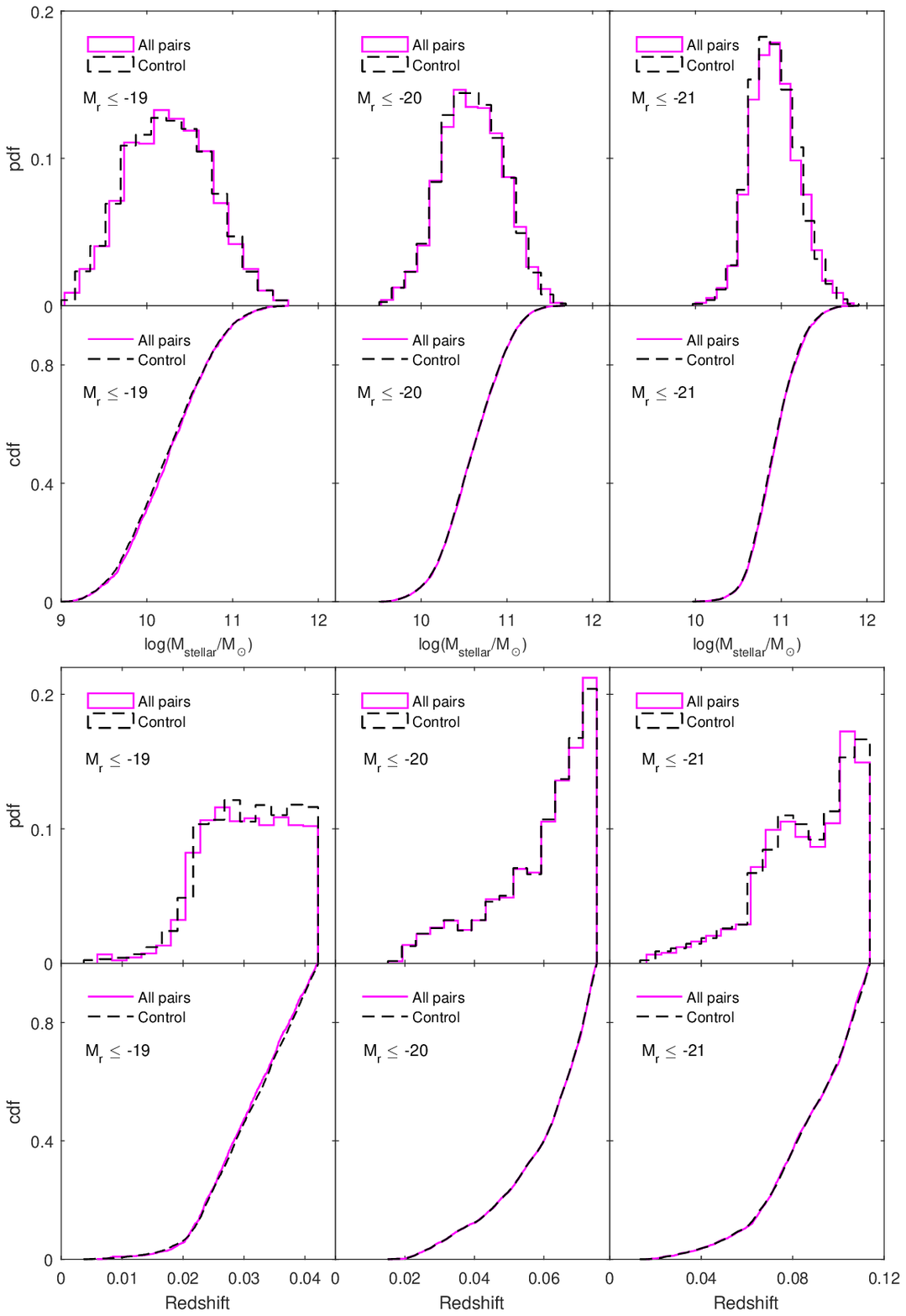}}}

\caption{The top two panels respectively shows PDF and CDF of
  $log(M_{stellar}/M_{sun})$ of all pairs and their corresponding
  control matched isolated galaxies in three absolute magnitude limits
  $M_r \leq -19$, $M_r \leq -20$ and $M_r \leq -21$. The bottom two
  panels show the same but for redshift.}
\label{control}
\end{figure*}

\begin{figure*}
\resizebox{15cm}{6cm}{\rotatebox{0}{\includegraphics{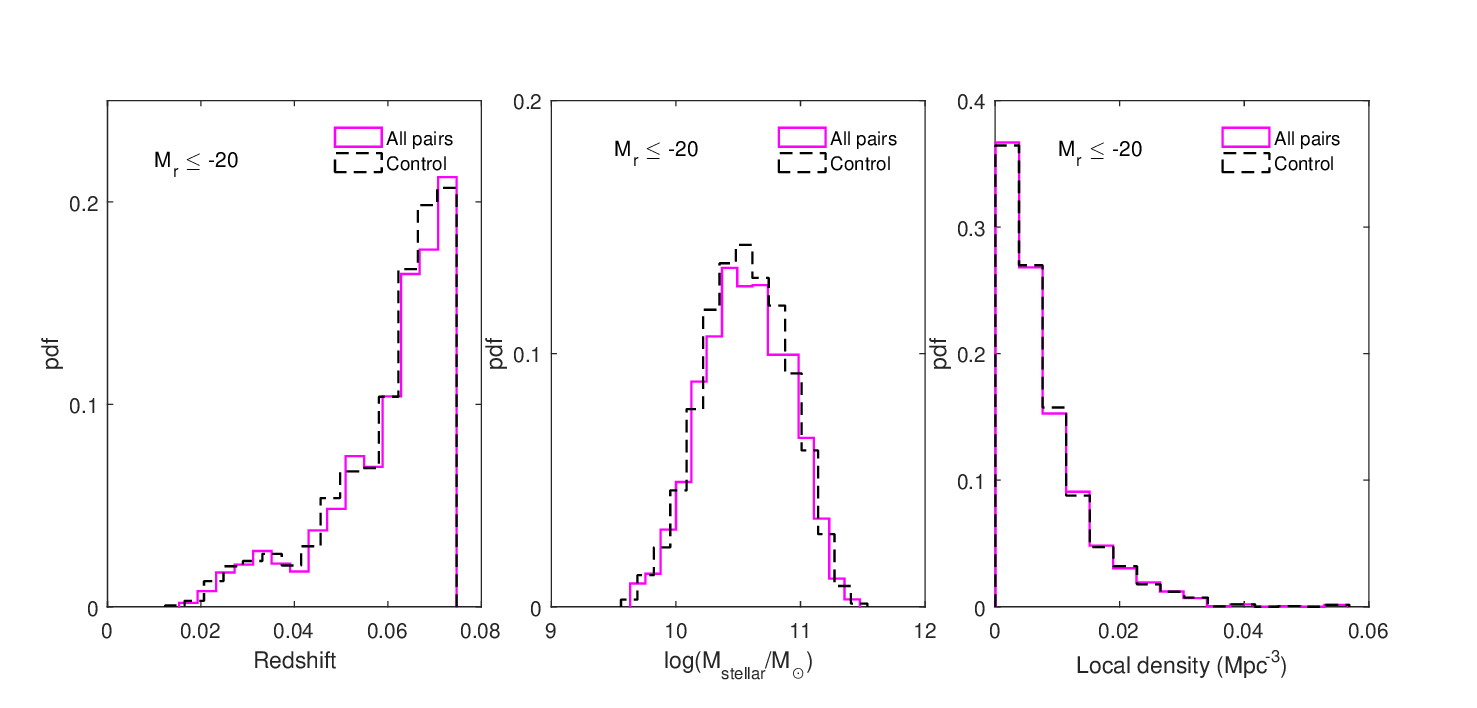}}}\
\caption*{}
\end{figure*}
\begin{figure*}
\resizebox{15cm}{6cm}{\rotatebox{0}{\includegraphics{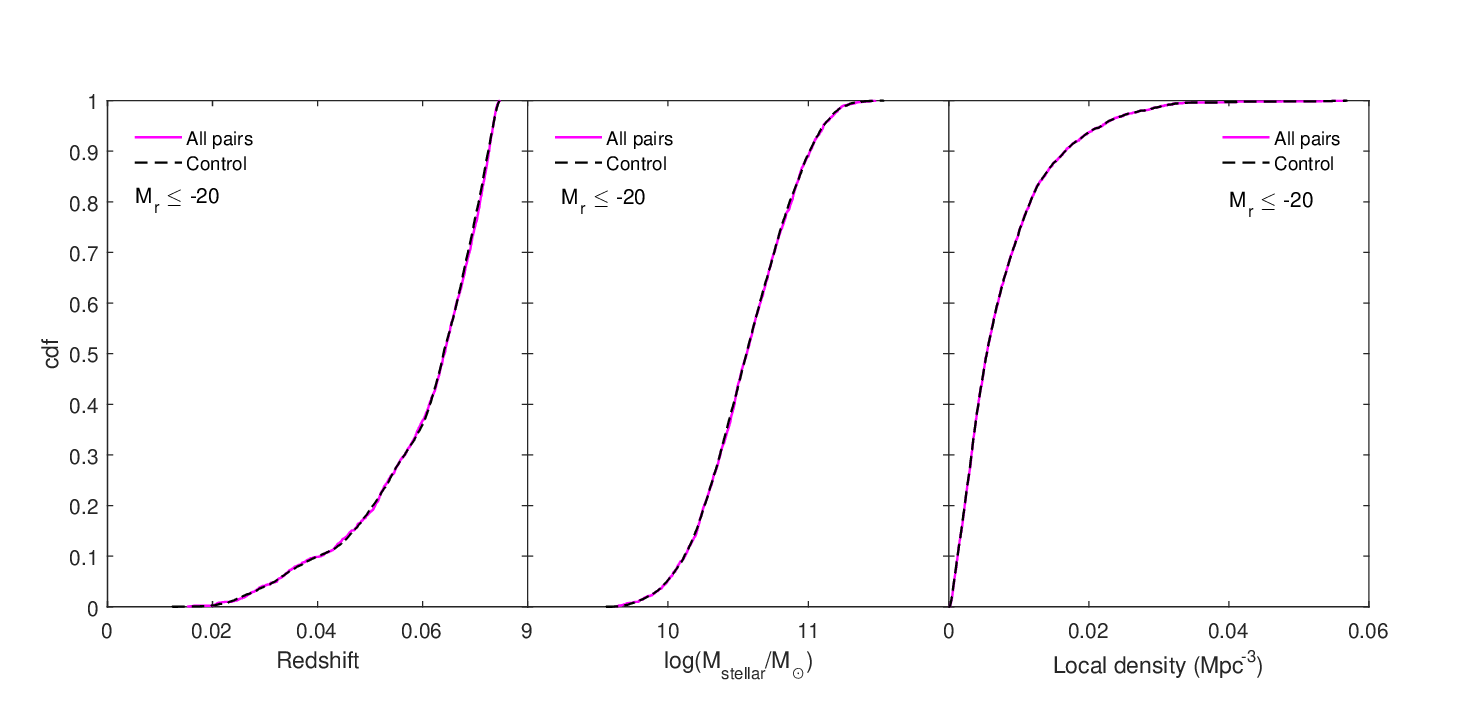}}}\
\caption{The top panels show the PDF of the redshift, stellar mass and
  local density of all the paired galaxies and their corresponding
  control matched isolated galaxies in the volume limited sample
  corresponding to the magnitude bin $M_r \leq -20$. The bottom panels
  show the respective CDFs.}
\label{control1}
\end{figure*}

\begin{table}
\centering
\begin{tabular}{|c|c|c|c|}
\hline
Volume limited sample & $M_r \leq -19 $ & $M_r \leq -20 $ & $M_r \leq -21 $\\
                      & ($z \leq 0.0422$) & ($z \leq 0.0752$) & ($z \leq 0.1137$)\\\hline
Number of galaxies & 21984 & 69456 & 85745\\
Number of pairs & 2581 & 5441 & 3039\\
Number of paired galaxies & 4032 & 9389 & 5679\\
Number of isolated galaxies & 17952 & 60067 & 80066\\
\hline
\end{tabular}
\caption{This table shows the total number of galaxies, the number of
  available pairs, the number of paired and isolated galaxies in the
  three volume limited samples considered in this work. The number of
  paired galaxies are not exactly twice the number of pairs as we
  allow a single galaxy to be part of multiple pairs provided they
  satisfy the pair selection criteria.}
\label{tab1}
\end{table}

\subsection{Identification of galaxy pairs}

\label{sec:findpairs}

We identify the galaxy pairs using the traditional method based on the
application of simultaneous cuts on the projected separation and the
velocity difference.

We calculate the projected separation ($r_p$) between any two galaxies
in the distribution using the following relation,
\begin{equation}
r_p= R\,\theta
\end{equation}
where $R$ represents the mean distance of the galaxy pair from the
observer given by,

\begin{equation} 
R=\frac{c}{2H_0}\bigg\{\int^{z_1}_0 \frac{dz}{E(z)}+\int^{z_2}_0 \frac{dz}{E(z)}\bigg\}.
\end{equation}

Here $c$ and $H_0$ carry their usual meaning and $E(z)$ in terms of $\Omega_{m0}$, $\Omega_{\Lambda0}$ and redshift $z$ is given by the following relation,

\begin{equation}
E(z)=\sqrt{\Omega_{m0}(1+z)^3+\Omega_{\Lambda0}}
\end{equation}

The angular separation $\theta$ between the two galaxies is,

\begin{equation}                        
\theta =\cos^{-1}\left[ \cos \delta_1 \,\cos \delta_2 \,\cos
  (\alpha_1-\alpha_2)+ \sin \delta_1 \,\sin \delta_2 \right].
\label{theta}
\end{equation}
Here $(\alpha_1, \delta_1)$ and $(\alpha_2, \delta_2)$ are the
equatorial co-ordinates of the two galaxies considered.

The difference between the rest frame Hubble velocities of the two galaxies is given by,

\begin{equation}                        
\Delta v = c \bigg|\frac{z_1}{1+z_1}-\frac{z_2}{1+z_2}\bigg|
\end{equation} 

In order to select the galaxy pairs, we impose simultaneous cuts on
the projected separation and the velocity difference of the two
galaxies under consideration. Any two galaxies are considered to form
a pair if their projected separation $r_p < 150$ kpc and the
rest-frame velocity difference $\Delta v<300$ km/s. It is known from
earlier studies that the pairs with larger separations are unlikely to
be interacting \citep{patton00, depropris07}.

An earlier work by \citet{scudder12b} shows that excluding the
galaxies with multiple companions do not alter their results. So we
allow a single galaxy to be part of multiple pairs provided they
satisfy the criteria imposed on $r_p$ and $\Delta v$.

The pair selection algorithm, when applied to the total 350536
galaxies from the contiguous region considered in our analysis, yields
a total 24756 galaxy pairs. We then cross-match the galaxies in the
volume limited sample for $M_r \leq -19$, $M_r \leq -20$, $M_r \leq
-21$ with the galaxies in the identified pairs. This provides us with
total 2581, 5441, 3039 galaxy pairs present in our volume limited
samples corresponding to the three magnitude bins listed in
\autoref{tab1}. We ensure that the matched galaxies in pairs must have
measurements of stellar mass, star formation rate and internal
reddening. We then impose another cut so as to only consider the pairs
with stellar mass ratio $\leq 10$. This restriction reduces the
available number of galaxy pairs to 2024, 5014 and 3002 in the three
volume limited samples considered in our work.

\begin{figure*}
\centering
\resizebox{9.5cm}{6.5cm}{\rotatebox{0}{\includegraphics{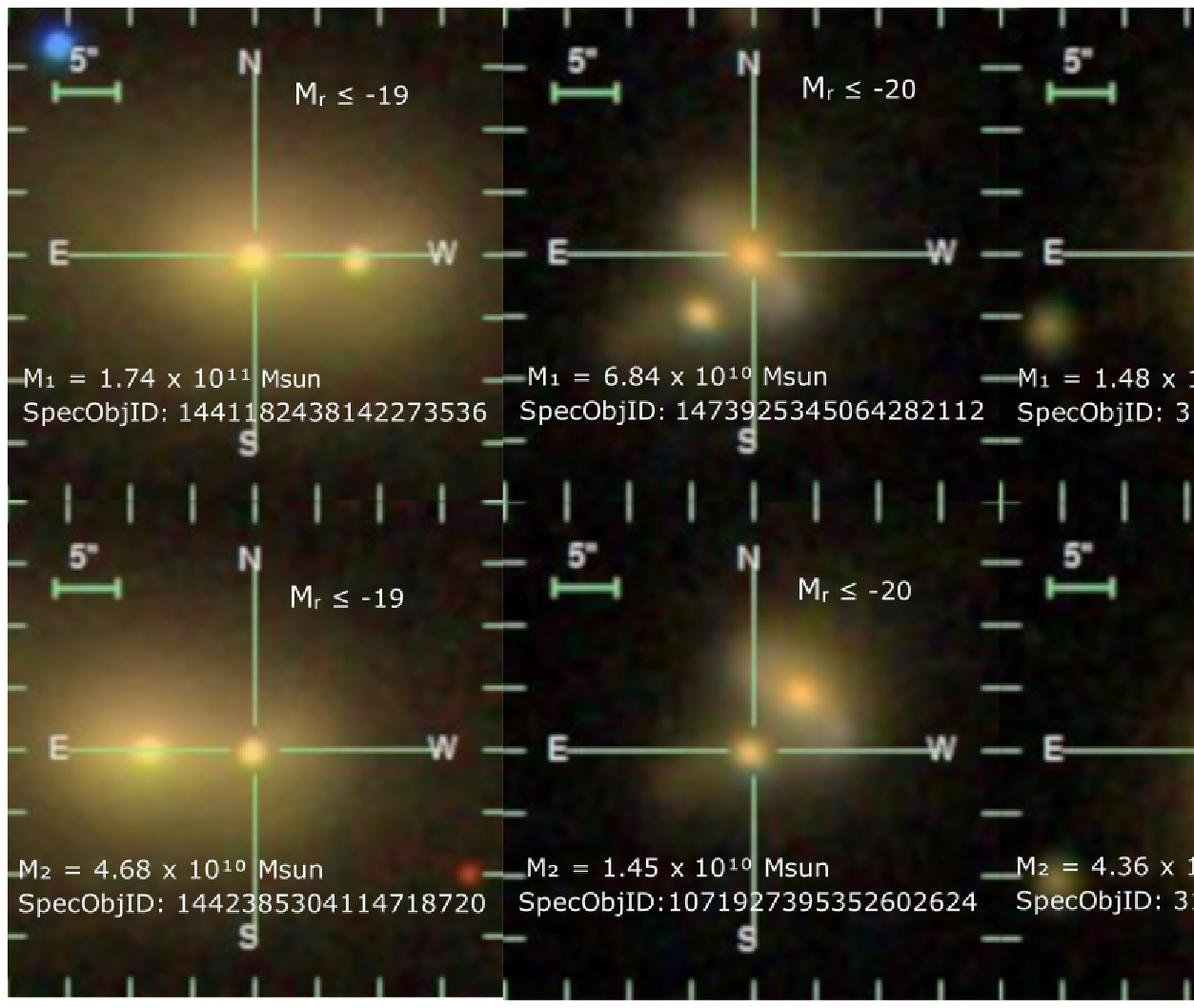}}}

\caption{This shows the image of one minor pair from each of the
  volume limited samples. The more massive and the less massive
  members in the minor pairs are marked in the top and bottom panels
  respectively. The stellar mass and SDSS $SpecObjID$ of the member
  galaxies are mentioned separately in each panel.}
\label{photo}
\end{figure*}

\subsection{SDSS fibre collision effect: culling pairs}
\label{sec:cullpairs}
It is important to take into account the spectroscopic incompleteness
due to the finite size of the SDSS fibres. The minimum separation of
the fibre centres in the SDSS is $55^{\prime\prime}$
\citep{strauss02}. Consequently, the companion galaxies closer than
$55^{\prime\prime}$ are preferentially missed. This leads to an
under-selection of the closer angular pairs. The galaxies within the
collision limit can still be observed if they lie in the overlapping
regions between adjacent plates. The ratio of spectroscopic to
photometric pairs decreases from $\sim 80\%$ at $>55^{\prime\prime}$
to $\sim 26\%$ at lower angular separation \citep{patton08}. This
incompleteness effect can be compensated by randomly culling $67.5\%$
of galaxies in pairs with the angular separation $>55^{\prime\prime}$
\citep{ellison08, patton11, scudder12b}.

Our three volume limited samples for $M_r \leq -19$, $M_r \leq -20$,
$M_r \leq -21$ respectively contains 2024, 5014 and 3002 galaxy pairs.
In a similar spirit to earlier works, we randomly exclude 67.5\% of
pairs which have their angular separation $ \theta  >
55^{\prime\prime}$ in all the three samples. The total number of
galaxy pairs available for further analysis after culling in the three
samples are 737, 2203 and 1600 respectively.  The galaxy pairs in the
three samples before and after culling are shown in
\autoref{fig:cullpair}.

In the three volume limited samples ($M_r \leq -19$, $M_r \leq -20$
and $M_r \leq -21$), there are respectively $17952$, $60067$, $80066$
(\autoref{tab1}) galaxies that have no identified pairs according to
the criteria applied in \autoref{sec:findpairs}. We term these
galaxies as isolated and use them to build our control sample as
described in the next subsection.

\begin{table*}
\centering
\begin{tabular}{c @{\vline} ccc @{\vline} cccccccc}
\hline
& \multicolumn{2}{c}{$D_{KS}$}& & \multicolumn{7}{c}{$D_{KS}(\alpha)$}\\

$M_r$ \hspace{2mm} & \hspace{2mm}Redshift & $log(M_{stellar}/M_{sun})$& & \hspace{5mm}99\% & 90\% & 80\% & 70\% & 60\% & 50\% & 40\% \\
\hline
$\leq -19$ \hspace{4mm}& \hspace{2mm}0.0267 & 0.0273 & & \hspace{5mm}0.0483 & 0.0363 & 0.0318 & 0.0289 & 0.0266 & 0.0247 & 0.0230 \\
$\leq -20$ \hspace{4mm}& \hspace{2mm}0.0116 & 0.0123 & & \hspace{5mm}0.0281 & 0.0211 & 0.0185 & 0.0168 & 0.0155 & 0.0144 & 0.0134 \\
$\leq -21$ \hspace{4mm}& \hspace{2mm}0.0181 & 0.0178 & & \hspace{5mm}0.0322 & 0.0242 & 0.0212 & 0.0192 & 0.0177 & 0.0164 & 0.0153\\
\hline

\end{tabular}
\caption{This table shows the Kolmogorov-Smirnov statistic $D_{KS}$
  for comparisons of redshift and $log(M_{stellar}/M_{sun})$ of all
  pairs and their corresponding control galaxies in three volume
  limited samples. The table also lists the critical values
  $D_{KS}(\alpha)$ above which the null hypothesis can be rejected at
  different confidence levels.}
\label{kstab}
\end{table*}

\begin{table*}
\centering
\begin{tabular}{c @{\vline} cccc @{\vline} ccccc}
\hline
& \multicolumn{3}{c}{$D_{KS}$}& & \multicolumn{5}{c}{$D_{KS}(\alpha)$}\\

$M_r$ \hspace{2mm} & \hspace{2mm}Redshift & $log(M_{stellar}/M_{sun})$& Local density & & \hspace{5mm}99\% & 90\% & 80\% & 70\% & 60\% \\
\hline
$\leq -20$ \hspace{4mm}& \hspace{2mm}0.0218 & 0.0223 & 0.0205 & & \hspace{5mm}0.0392 & 0.0295 & 0.0258 & 0.0235 & 0.0216  \\
\hline

\end{tabular}
\caption{This table shows the Kolmogorov-Smirnov statistic $D_{KS}$
  for comparisons of redshift, $log(M_{stellar}/M_{sun})$ and local
  density of all pairs and their corresponding control galaxies in the
  volume limited sample corresponding to the magnitude bin $M_r \leq
  -20$. The critical values $D_{KS}(\alpha)$ above which the null
  hypothesis can be rejected at different confidence levels are also
  listed in the same table.}
\label{kstab1}
\end{table*}


\begin{table*}
\centering
\begin{tabular}{c @{\vline} ccc @{\vline} ccccccccc}
\hline
& \multicolumn{2}{c}{$D_{KS}$} & & \multicolumn{8}{c}{$D_{KS}(\alpha)$}\\

$M_r$\hspace{2mm} & \hspace{2mm}SFR & $(u-r)$ & & \hspace{2mm}99\% & 90\% & 80\% & 70\% & 60\% & 50\% & 40\% & 30\% \\
\hline
$\leq -19$\hspace{2mm} & \hspace{2mm}0.1855 & 0.1944 & & \hspace{2mm}0.0686 & 0.0516 & 0.0452 & 0.0411 & 0.0378 & 0.0351 & 0.0327 & 0.0289\\

$\leq -20$\hspace{2mm} & \hspace{2mm}0.1412 & 0.1481 & & \hspace{2mm}0.0464 & 0.0349 & 0.0306 & 0.0277 & 0.0256 & 0.0237 & 0.0221 & 0.0202\\

$\leq -21$\hspace{2mm} & \hspace{2mm}0.1773 & 0.1908 & & \hspace{2mm}0.0688 & 0.0518 & 0.0454 & 0.0412 & 0.0379 & 0.0352 & 0.0328 & 0.0289\\

\hline

\end{tabular}
\caption{This table shows the Kolmogorov-Smirnov statistic $D_{KS}$
  for comparisons of star formation rate (SFR) and $(u-r)$ colour of
  minor pairs and their control galaxies in the three volume limited
  samples. The control galaxies are matched in stellar mass and
  redshift. The table also lists the critical values $D_{KS}(\alpha)$
  above which the null hypothesis can be rejected at different
  confidence levels.}

\label{tab5}
\end{table*}

\begin{table}
\centering
\begin{tabular}{|c|c|c|c|}
 & $M_r \leq -19$ & $M_r \leq -20$ & $M_r \leq -21$ \\
\hline
Galaxies in minor pairs & 675 & 1479 & 671 \\
after control matching & & &\\
\hline
Elliptical & 147 & 218 & 123\\
Spiral & 185 & 402 & 160\\
Uncertain morphology & 308 & 796 & 351\\
Morphology not available & 35 & 63 & 37\\
\hline
Dominant bulge & 8 & 13 & 6\\
Bar & 24 & 56 & 49\\
AGN & 72 & 98 & 38\\
\hline
Minor pairs & 350 & 756 & 338\\
after control matching & & &\\
\hline
Spiral - Spiral & 37 & 66 & 30\\
Elliptical - Elliptical & 22 & 13 & 12 \\
Spiral - Elliptical & 34 & 51 & 25\\
\hline
\end{tabular}
\caption{This table shows number of the minor paired galaxies that are
  classified as elliptical, spiral, uncertain morphology and those
  with the presence of bar, dominant bulge or AGN activity. It also
  shows the number of minor pairs with Spiral-Spiral,
  Elliptical-Elliptical and Spiral-Elliptical combination. 
}

\label{morpho}
\end{table}

\begin{table*}
\centering
\begin{tabular}{c @{\vline} ccc @{\vline} ccccccccc}
\hline
& \multicolumn{2}{c}{$D_{KS}$} & & \multicolumn{8}{c}{$D_{KS}(\alpha)$}\\

$M_r$\hspace{2mm} & \hspace{2mm}SFR & $(u-r)$ & & \hspace{2mm}99\% & 90\% & 80\% & 70\% & 60\% & 50\% & 40\% & 30\% \\
\hline
$\leq -20$\hspace{2mm} & \hspace{2mm}0.1011 & 0.0843 & & \hspace{2mm}0.0702 & 0.0528 & 0.0463 & 0.0420 & 0.0387 & 0.0359 & 0.0335 & 0.0313\\
\end{tabular}

\caption{ Same as \autoref{tab5} but for the volume limited sample
  corresponding to the magnitude bin $M_r \leq -20$ and the control
  galaxies that are matched in stellar mass, redshift and local
  density.}
\label{tabextra}
\end{table*}

\subsection{Building control sample}
\label{sec:csample}
The physical properties of interacting galaxies should be compared
against carefully designed control sample of non-interacting galaxies.
The colour and SFR of galaxies depends on their stellar mass. So it is
crucial to ensure that the distributions of stellar mass for the pairs
and control samples are statistically indistinguishable. The colour
and the star formation activity of galaxies are also known to depend
on the redshift. The redshift dependent selection effects can not be
eliminated completely even in a volume limited sample. So we also
decide to match the redshift distributions of the paired galaxies and
control sample of isolated galaxies.
 
After correcting for the fibre collision effect by culling galaxy
pairs as described in \autoref{sec:cullpairs}, we have total 737, 2203
and 1600 pairs in the three volume limited samples.  We adopt a
strategy similar to \citet{ellison08} for building the control
sample. We build the control sample of the isolated galaxies by
simultaneously matching their stellar mass and redshift with that of
the paired galaxies. For each paired galaxy, we pick $5$ unique
isolated galaxies matched in stellar mass and redshift. We match the
paired galaxies and their controls within 0.085, 0.050, 0.061 dex in
stellar mass and 0.0065, 0.0010, 0.0030 in redshift for the three
samples corresponding to the magnitude bins $M_r \leq -19$, $M_r \leq
-20$, $M_r \leq -21$ respectively. We match every paired galaxy to
their controls and then perform a Kolmogorov-Smirnov (KS) test on the
stellar mass and redshift distributions.  The PDFs and CDFs of the
paired and control matched galaxies are shown in
\autoref{control}. The results for the corresponding KS tests are
tabulated in \autoref{kstab}. The control samples are accepted only
when their stellar mass and redshift distributions are consistent with
that of the paired galaxies at a level of at least $30\%$ KS
probability for $M_r \leq -20$ and $60\%$ KS probability for the
remaining two magnitude bins. This implies that the null hypothesis
can be rejected at $\leq 30\%$ confidence level (\autoref{kstab}) for
$M_r \leq -20$ and $\leq 60\%$ confidence level (\autoref{kstab}) for
$M_r \leq -19$ and $M_r \leq -21$. This ensures that the galaxies in
the pair and control samples are highly likely to be drawn from the
same parent redshift and stellar mass distributions. The two magnitude
bins $M_r \leq -19$ and $M_r \leq -21$ contain relatively smaller
number of pairs as compared to the magnitude bin $M_r \leq -20$. So a
somewhat higher threshold for the confidence level was used to prepare
the control samples in these two magnitude bins.

\begin{figure*}
\resizebox{15cm}{6cm}{\rotatebox{0}{\includegraphics{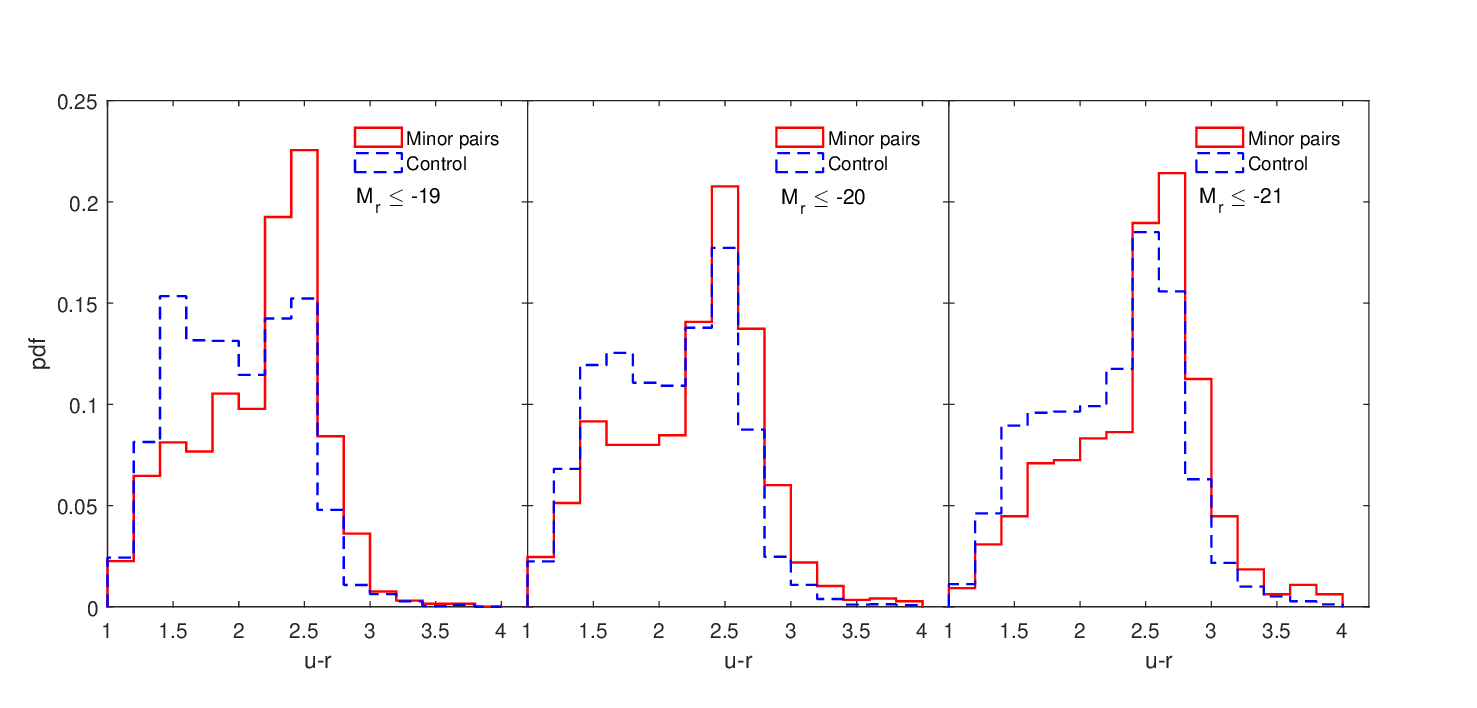}}}\
\caption*{}
\end{figure*}

\begin{figure*}
\resizebox{15cm}{6cm}{\rotatebox{0}{\includegraphics{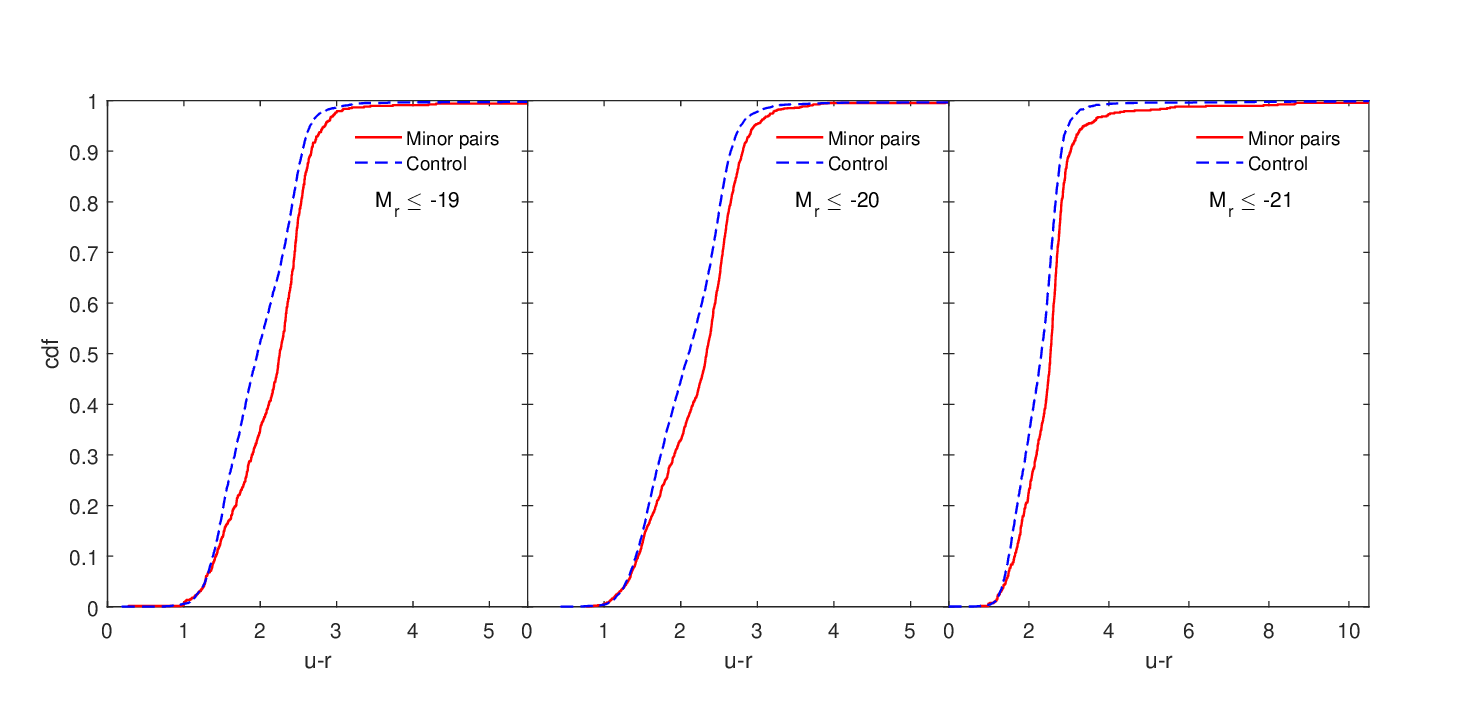}}}\
\caption{The top and bottom panels respectively show the PDF and CDF
  of $(u-r)$ colour of all the minor pairs and their corresponding
  control matched isolated galaxies in the three volume limited
  samples.}
\label{fig3}
\end{figure*}
\begin{figure*}
\resizebox{15cm}{6cm}{\rotatebox{0}{\includegraphics{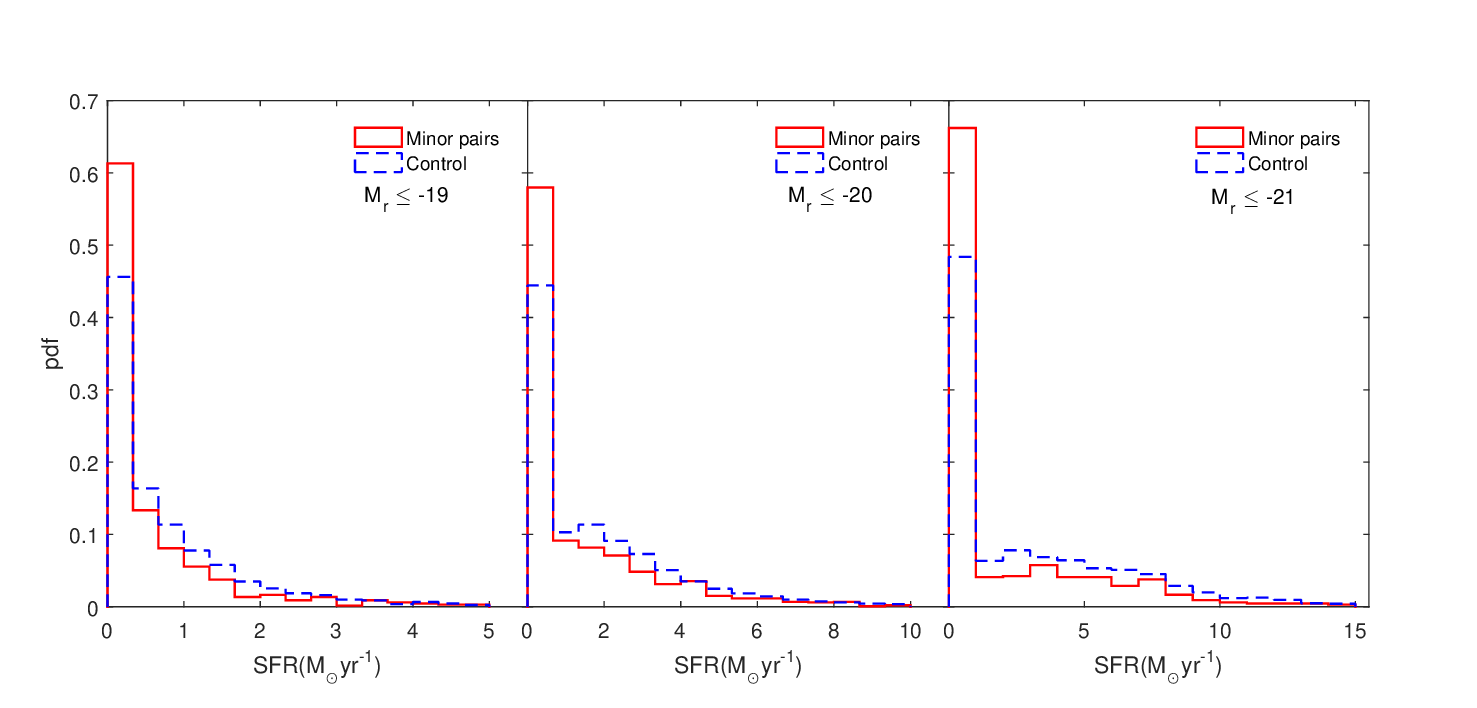}}}
\caption*{}
\end{figure*}
\begin{figure*}
\resizebox{15cm}{6cm}{\rotatebox{0}{\includegraphics{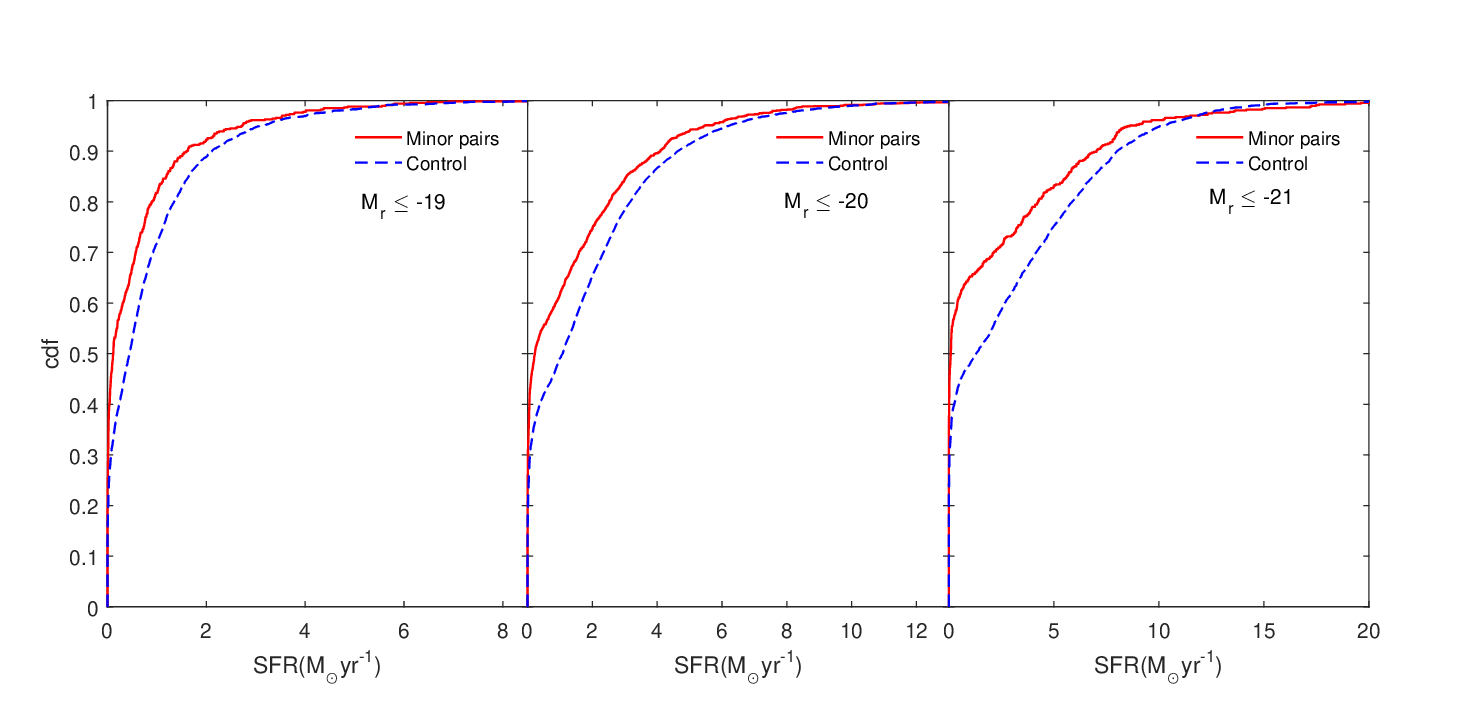}}}

\caption{Same as \autoref{fig3}, but for the star formation rate.}
\label{fig4}
\end{figure*}
All the paired galaxies in our volume limited samples have
measurements of stellar mass and redshift. But the condition that
there should be at least $5$ control galaxies for each paired galaxy
reduces the number of available galaxy pairs. After the control
matching, we have 737, 2159 and 1592 galaxy pairs in the three volume
limited samples corresponding to the magnitude bins $M_r \leq -19$,
$M_r \leq -20$, $M_r \leq -21$ respectively. These galaxy pairs are
formed by a total 1363, 4032, 3074 galaxies respectively. The control
sample of these paired galaxies consists of a total 6815, 20160, 15370
isolated galaxies respectively. The control matching in stellar mass
and redshift eliminate most of the biases that can plague a comparison
between the two samples \citep{perez09b}.

We define minor pairs as those which have their stellar mass ratio in
the range $3 \leq \frac{M_1}{M_2} \leq 10$.  We find that there are
350, 756, 338 minor pairs after control matching in the three volume
limited samples corresponding to the magnitude bins $M_r \leq -19$,
$M_r \leq -20$, $M_r \leq -21$ respectively. These minor pairs are
respectively formed by 675, 1479 and 671 galaxies. We note that the
minimum projected separation between the galaxies in the minor pairs
in all the three volume limited samples are $\sim 4$ kpc. We show the
image of one representative minor pair from each of the three volume
limited samples in \autoref{photo}. The $SpecObjID$ and the stellar
mass of the member galaxies are provided in each panel of
\autoref{photo}.

The environment plays a crucial role in determining the galaxy
properties. The paired galaxies may preferentially reside in the high
density environments. This may affect any comparison between the
colour and SFR of the paired galaxies and the isolated galaxies in our
sample. To account this, we also match the local density of the paired
and isolated galaxies besides matching their stellar mass and
redshift. However, this significantly reduces the number of available
minor pairs in our samples. So we simultaneously match the stellar
mass, redshift and local density of paired and their controls only for
the volume limited sample corresponding to the magnitude bin $M_r \leq
-20$. Here the paired galaxies and their controls are matched within
0.005 in redshift, 0.08 dex in stellar mass and 0.001 in local density
(Mpc$^{-3}$). The PDFs and CDFs of the paired and control matched
galaxies are shown in \autoref{control1}. The results for the
corresponding KS tests are tabulated in \autoref{kstab1}.  We get a
total 1076 pairs after control matching out of which 328 are minor
pairs.

We estimate the local number density at the location of each galaxies
by using the $k^{th}$ nearest neighbour method
\citep{casertano85}. The local number density is defined as,
\begin{eqnarray}
{\eta}_k = \frac{k-1}{\frac{4}{3}\pi r_k^3}  
\label{eqn:knn}
\end{eqnarray}  
Here $r_k$ is the distance between the galaxy and its $k^{th}$
nearest neighbour. We choose $k=5$ for the present analysis.\\

\begin{figure*}
\resizebox{15cm}{6cm}{\rotatebox{0}{\includegraphics{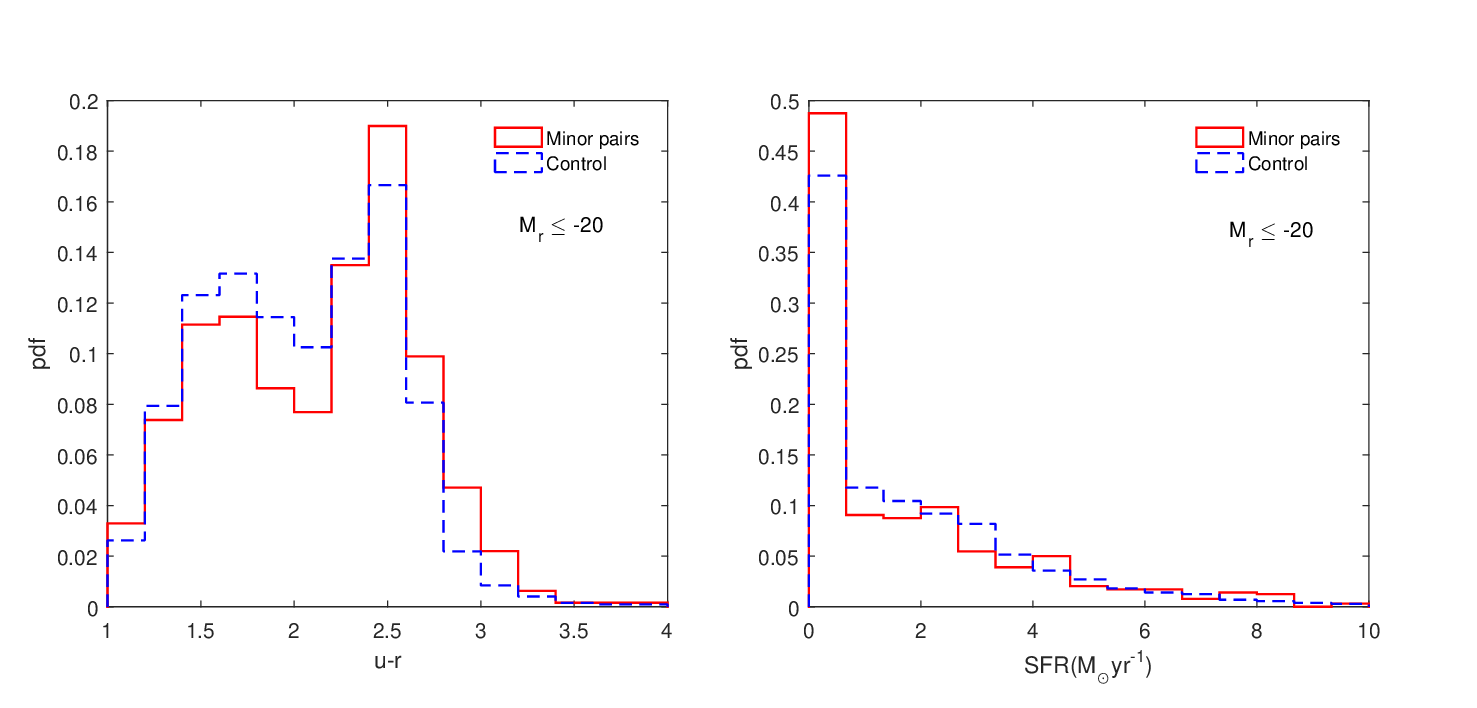}}}\
\caption*{}
\end{figure*}
\begin{figure*}
\resizebox{15cm}{6cm}{\rotatebox{0}{\includegraphics{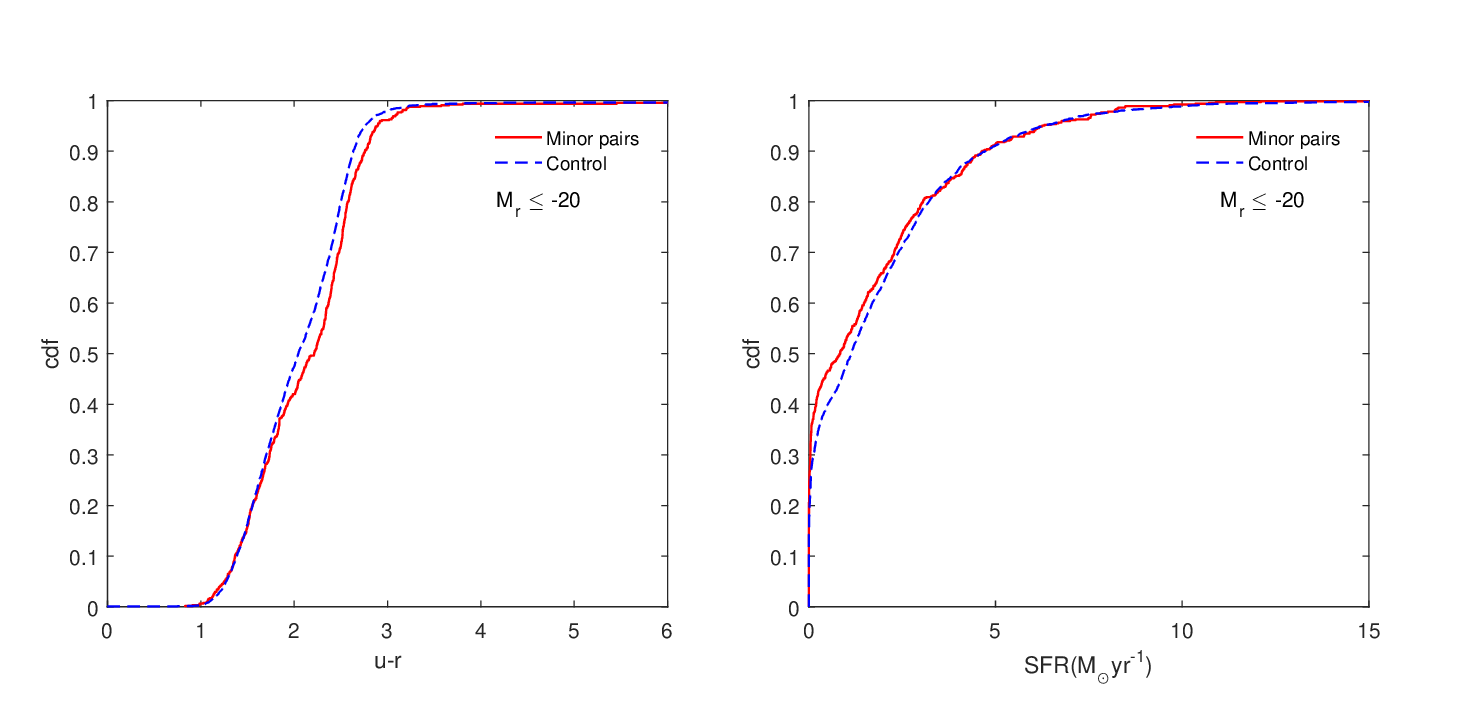}}}\
\caption{The top left panel of this figure compares the PDF of $(u-r)$
  colour for the galaxies in minor pairs and their control galaxies
  matched in stellar mass, local density and redshift. The top right
  panel compares the same for SFR. The bottom two panels compares the
  respective CDFs. This figure shows only the results for the volume
  limited sample corresponding to the magnitude bin $M_r \leq -20$. }
\label{figextra}
\end{figure*}

\subsection{Colour and SFR offsets}
\label{sec:off}
The ongoing interaction between the galaxies in pair can influence
their colour and star formation. The paired galaxies would be bluer
compared to the control matched isolated galaxies when SFR is enhanced
due to the tidal interaction. However, the colour of a galaxy pair can
also become redder than the colour of its control matched isolated
galaxies. This could happen due to the fact that paired galaxies
reside in high density regions than the regions occupied by control
sample of isolated galaxies. This reddening of colour is therefore
also true for pairs in which interaction between the two members has
not started and are in the stage of pre-interaction
\citep{patton11}. Such a change in colour of the galaxies in pair and
their corresponding control galaxies would not be detected
individually and only an overall change could be detected by comparing
the colour distributions of the entire pair and control samples.

We compute the colour offset of every paired galaxy in our sample with
respect its control galaxies. \citet{patton11} define the color offset
of individual paired galaxy as difference of its colour and the mean
colour of its 5 associated control galaxies. Similarly, the colour
offset of each control galaxy is defined as the difference of its
colour and the mean colour of the remaining 4 control
galaxies. Following the strategy, we compute the colour offset of each
galaxy in minor pairs and the corresponding control galaxies in each
of the three volume limited samples considered in this work. We then
compute the difference between the colour offsets of all galaxies in
minor pairs and their control galaxies at each projected separation
($r_p$). We estimate the cumulative mean of this colour offset
difference $\Delta$$(u-r$ offset) as a function of the projected
separation. It maybe noted that if colour of paired and control
galaxies represent subsets of the same colour distribution, then the
average value of colour offset difference is expected to be zero
\citep{patton11}.

We also calculate the star formation rate (SFR) offset of each galaxy
in minor pairs and their corresponding control galaxies in all the
volume limited samples in a similar manner. The cumulative mean of the
SFR offset difference $\Delta$($SFR$ offset) is estimated as a
function of the projected separation. The $\Delta$($SFR$ offset) is
expected to be positive for SFR enhancement, negative for suppression
in SFR and zero when there are no differences in the SFR of the paired
galaxies and their controls.

\begin{figure*}
\resizebox{15cm}{6cm}{\rotatebox{0}{\includegraphics{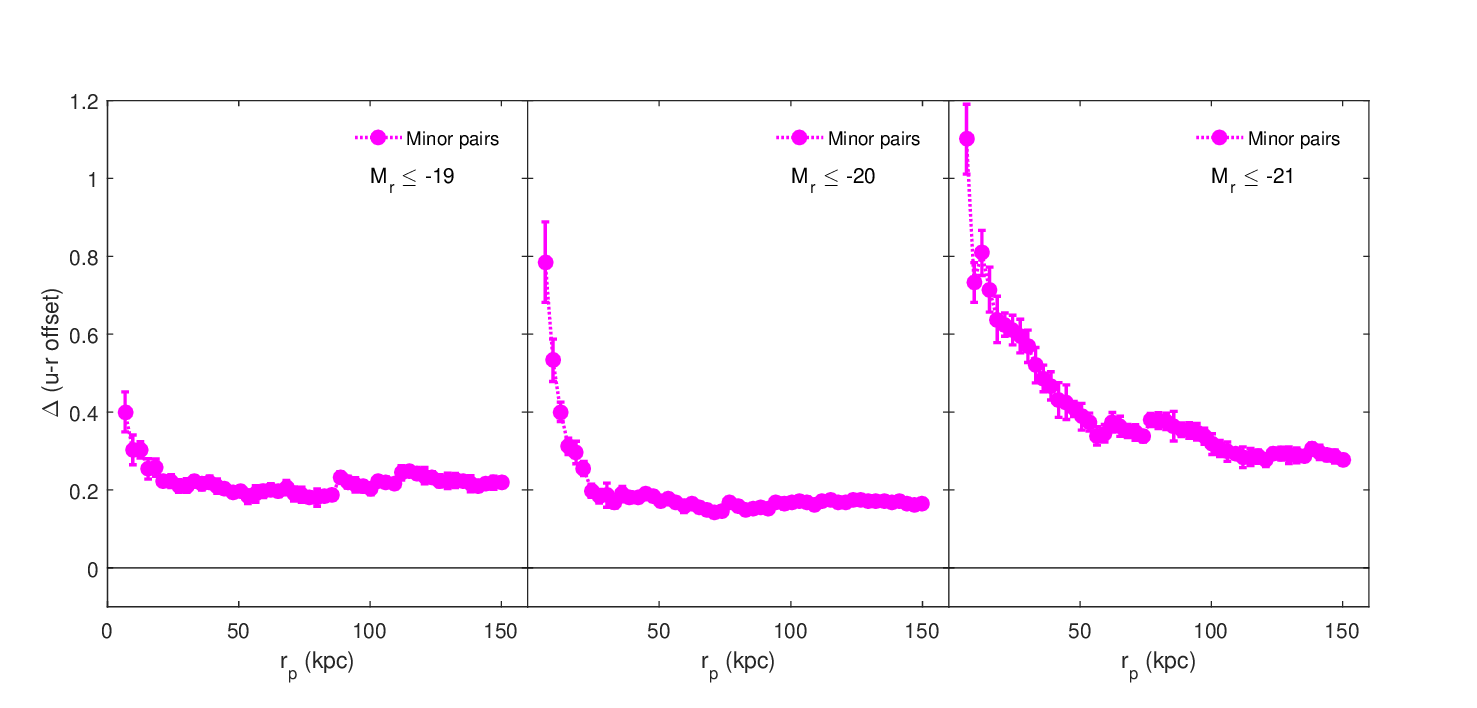}}}\
\resizebox{15cm}{6cm}{\rotatebox{0}{\includegraphics{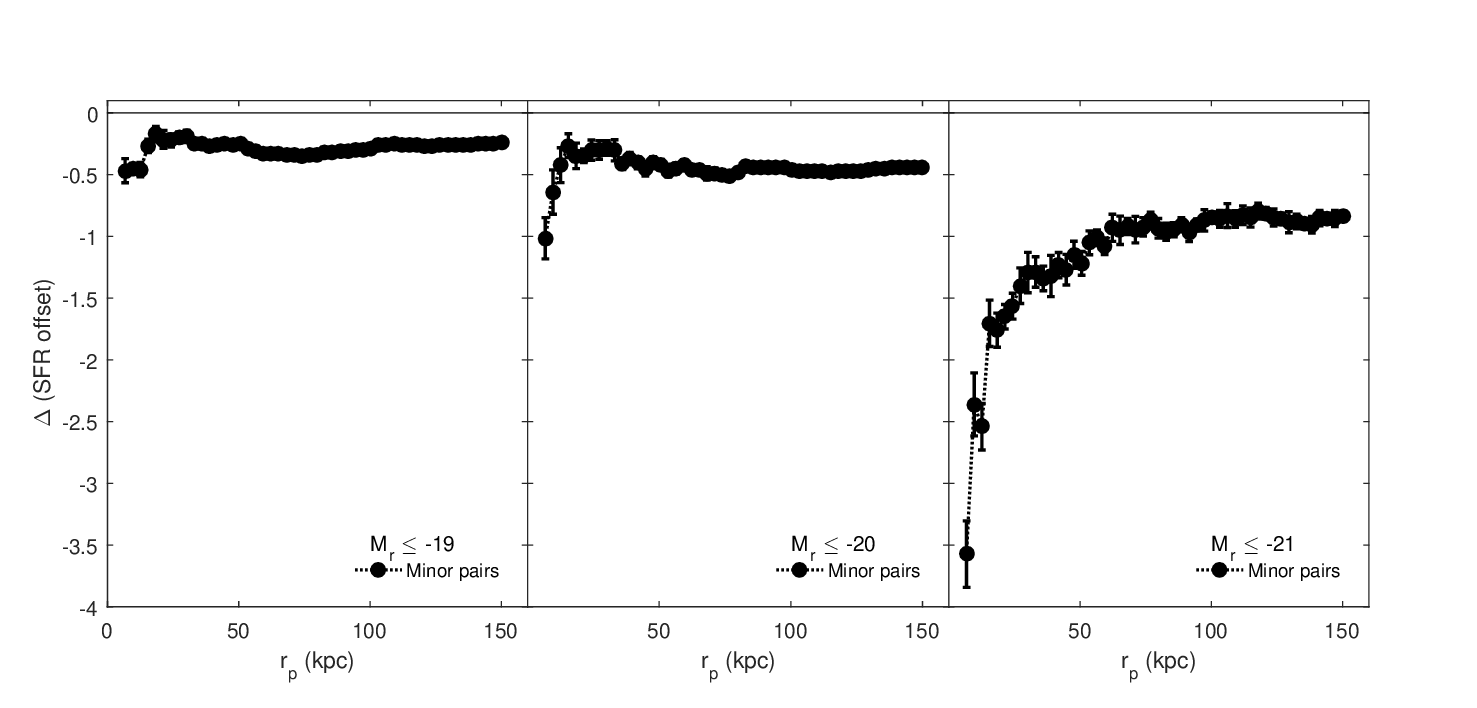}}}\
\caption{The top and bottom panels respectively show the
  $\Delta$$(u-r$ offset) and $\Delta$($SFR$ offset) as a function of
  the projected separation for the three volume limited samples. Here
  the control galaxies are simultaneously matched in stellar mass and
  redshift. The $1-\sigma$ errorbars shown here are estimated using 10
  Jackknife samples.}
\label{fig6}
\end{figure*}

\section{Results}
\subsection{Comparing SFR and $(u-r)$ colour distributions in the minor pairs and the isolated galaxies after matching the stellar mass and redshift}
\label{res1}
We would like to understand the role of minor interactions on the star
formation rate and colour of galaxies in the present Universe. It is
important to test whether the galaxies in minor pair and their control
sample of isolated galaxies have any differences in their colour and
SFR distributions. We compare the $(u-r)$ colour distribution of the
galaxies in minor pairs and their control samples in three volume
limited samples in the top three panels of \autoref{fig3}. The
galaxies can be divided into blue and red classes by employing a cut
in their $(u-r)$ colour. The galaxies with $(u-r)<2.22$ and
$(u-r)>2.22$ can be labelled as blue and red respectively
\citep{strateva01}. We find more red galaxies in the minor pairs
compared to their control samples in each of the volume limited
samples. Besides, the control samples of the isolated galaxies contain
a relatively larger number of blue galaxies than the samples of minor
pairs. We also compare the CDFs of the $(u-r)$ colour for the paired
and control galaxies in each volume limited sample in the bottom
panels of \autoref{fig3}. We perform a KS test to check if the
distributions of $(u-r)$ colour in minor pairs and their control
samples are statistically different in a significant manner. The
results of the KS test are tabulated in \autoref{tab5} which clearly
shows that the null hypothesis can be rejected at $>99\%$ confidence
level for each volume limited sample. So the $(u-r)$ colour of the
minor pairs are significantly different from the galaxies in their
control samples.

We perform a similar analysis with the SFR of the galaxies in the
minor pairs and their controls. The PDFs and CDFs for the paired and
isolated galaxies are compared in the top and bottom panels of
\autoref{fig4}. The top panels of \autoref{fig4} show that the minor
pairs contain a larger number of low star-forming galaxies compared
to their control samples in each volume limited sample. Noticeably,
the number of high star-forming galaxies in the minor pairs is lower
than the respective control samples of the isolated galaxies. The
results of the KS test for SFR distributions are also tabulated in
\autoref{tab5}. We find that the SFR of the minor pairs and their
controls are significantly different, and the null hypothesis can be
rejected at $>99\%$ confidence level.

\begin{figure*}

\resizebox{15cm}{6cm}{\rotatebox{0}{\includegraphics{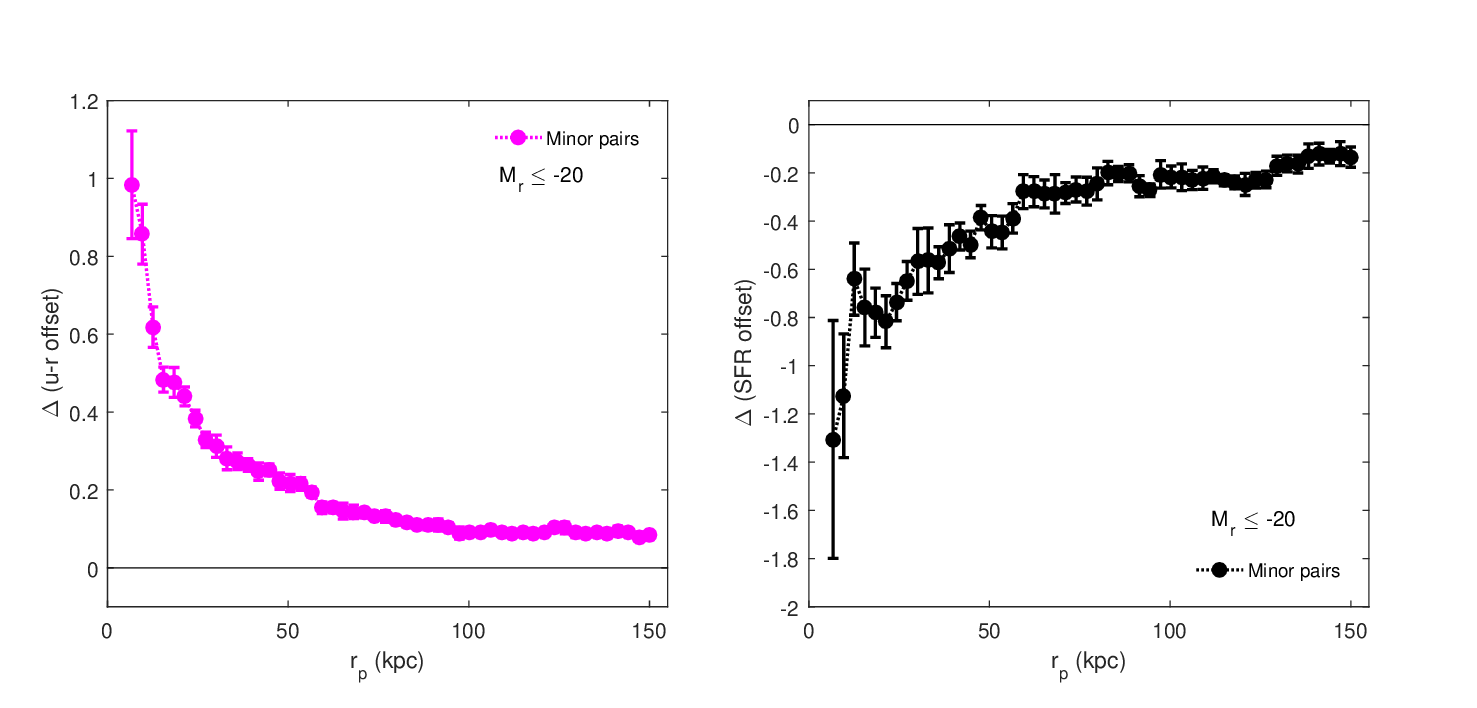}}}\
\caption{The left and right panels respectively show the
  $\Delta$$(u-r$ offset) and $\Delta$($SFR$ offset) as a function of
  the projected separation $(r_p)$ for the volume limited sample
  corresponding to the magnitude bin $M_r \leq -20$. Here, we match
  the control galaxies simultaneously in stellar mass, local density
  and redshift.  The $1-\sigma$ errorbars shown here are estimated
  using 10 Jackknife samples.}
\label{figadd2}
\end{figure*}

We show the cumulative mean of the $\Delta$$(u-r$ offset) as a
function of projected separation for the three volume limited samples
in the top panels of \autoref{fig6}. The results show that the
$\Delta$$(u-r$ offset) is positive at all pair separations upto 150
kpc. The $\Delta$$(u-r$ offset) tends to decrease with increasing pair
separation but nearly plateaus beyond 50 kpc in all three volume
limited samples. The magnitude of $\Delta$$(u-r$ offset) is relatively
higher in the brighter samples. The cumulative mean of the
$\Delta$($SFR$ offset) as a function of projected separation are shown
for all three volume limited samples in the bottom panels of
\autoref{fig6}. The $\Delta$($SFR$ offset) is negative throughout the
entire pair separation range. The $\Delta$($SFR$ offset) increases
with increasing pair separation upto a distance of $\sim 50$ kpc and
nearly plateaus out beyond this pair separation. We also note that the
magnitude of $\Delta$($SFR$ offset) is systematically lower for
brighter samples. A higher $\Delta$$(u-r$ offset) and lower
$\Delta$($SFR$ offset) for brighter samples indicate a luminosity
dependent quenching. Such trends may arise because brighter pairs
reside in denser environments where the galaxies are mostly red.

A positive $\Delta$$(u-r$ offset) indicates that, on average, galaxies
in minor pairs are redder than galaxies without a close companion. A
negative $\Delta$($SFR$ offset) corresponds to the suppression of star
formation in the minor pairs. So the trends observed in $\Delta$$(u-r$
offset) and $\Delta$($SFR$ offset) are consistent. Our results
indicate that the minor interactions may initiate quenching in
galaxies in the present universe. However, it is also important to
note that the environment may play an important role in such
quenching.

The control samples are matched in stellar mass and redshift, which
ensures that any differences in the colour and SFR distributions of
the minor pairs and their controls do not arise due to differences in
their stellar mass and redshift distributions. The observed
differences in colour and SFR may be caused by the differences in
their environments. It can be tested by matching the environment of
the controls before the comparison.

\subsection{Morphology of the galaxies in the minor pairs}

It is well known that the tidal interactions may cause gas loss
through AGN or shock-driven winds \citep{murray05, springel05}, induce
bar quenching \citep{haywood16} and morphological quenching
\citep{martig09}. We want to check the bar, dominant bulge and AGN
occurrences in the minor pairs. We cross-match the minor pairs with
the galaxies in Galaxy Zoo \citep{lintott08} and Galaxy Zoo2
\citep{willett} to reveal the morphological properties of the member
galaxies in minor pairs. The morphological information of the galaxies
in the minor pairs is listed in \autoref{morpho}. It is worthwhile to
mention that the morphological information is not available for $\sim
50\%$ minor pairs in our sample. We examine only the minor pairs for
which the morphological information is available. The Spiral-Spiral
and Spiral-Elliptical combinations in minor pairs are more prevalent
than the Elliptical-Elliptical combination. Only $4\%-7\%$ galaxies in
the minor pairs host a bar. A dominant bulge is observed in only $\sim
1\%$ of the galaxies in the minor pairs. $5\%-10\%$ galaxies in the
minor pairs show the AGN activity. We observe that only a small
percentage of the galaxies in the minor pairs host bar, dominant bulge
or AGN.  We do not compare the SFR in minor pairs to that with the
galaxies hosting bar, dominant bulge or AGN. Since the galaxies with
AGN, bar and dominant bulge are known to initiate quenching in
galaxies, a dominance of such galaxies in minor pairs would indicate
that the observed quenching in minor pairs are primarily driven by
these factors. We do not observe any such trends in our analysis.

\subsection{Comparing SFR and $(u-r)$ colour distributions in the minor pairs and the isolated galaxies after matching the stellar mass, redshift and local density}
The quenching in minor pairs may also arise due to the
environment. The paired galaxies preferentially reside in the denser
environments that host the redder galaxies. So the overabundance of
red and low star-forming galaxies in the minor pairs compared to their
control sample of isolated galaxies may not be caused by the
interactions alone. One needs to compare these properties by also
matching the environment of the control galaxies. We prepare the
control samples by simultaneously matching the stellar mass, redshift
and the local density of the isolated galaxies. We find that such
controls can be prepared only for a single volume limited sample
($M_r<-20$) due to the smaller number of minor pairs present in our
samples.

We compare the PDFs of $(u-r)$ colour and SFR in the minor pairs and
their control galaxies in the top two panels of \autoref{figextra}.
Interestingly, the minor pairs still host more red and
low star-forming galaxies compared to their controls. The differences
in their PDFs are smaller than those observed in \autoref{fig3} and
\autoref{fig4} though. The respective CDFs are compared in the two
bottom panels of \autoref{figextra}. We perform KS tests to quantify
the differences in the distributions of $(u-r)$ colour and SFR in the
minor pairs and their control galaxies. The results of the KS tests
are listed in \autoref{tabextra} which clearly show that the null
hypothesis can still be rejected at $>99\%$ confidence level.

We show the cumulative mean of the $\Delta$$(u-r$ offset) and
$\Delta$($SFR$ offset) as a function of pair separation in the two
panels of \autoref{figadd2}. The \autoref{figadd2} exhibits the same
trends as observed in the \autoref{fig6}. We find that $\Delta$$(u-r$
offset) decreases, and $\Delta$($SFR$ offset) increases with the
increasing pair separation. The dependence of the quenching on the
pair separation is noticeable at least upto $\sim 50$ kpc. The degree
of suppression of the star formation in minor pairs remains nearly
unchanged at greater separations.

The analysis in this subsection implies that the environment has some
role in quenching the star formation in minor pairs. Nevertheless, the
signature of quenching in minor pairs persists even after controlling
their environment, stellar mass and redshift. It indicates that the
minor interaction induces quenching in galaxies in the present
Universe.

\begin{figure*}

\resizebox{15cm}{6cm}{\rotatebox{0}{\includegraphics{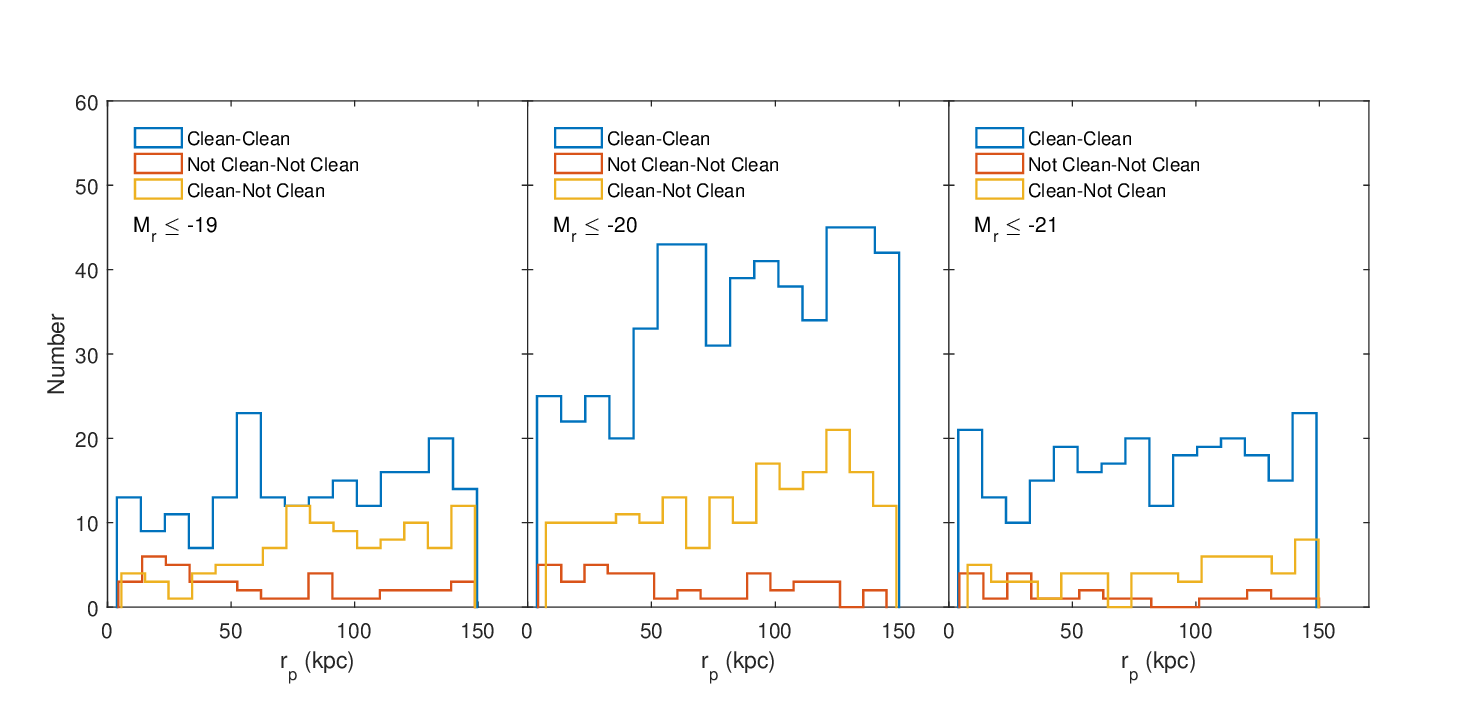}}}\
\caption{The three panels of this figure show the number of minor
  pairs with clean/unclean photometry as a function of the projected
  separations in the three volume limited samples.}
\label{photo}
\end{figure*}

\begin{figure*}

\resizebox{15cm}{6cm}{\rotatebox{0}{\includegraphics{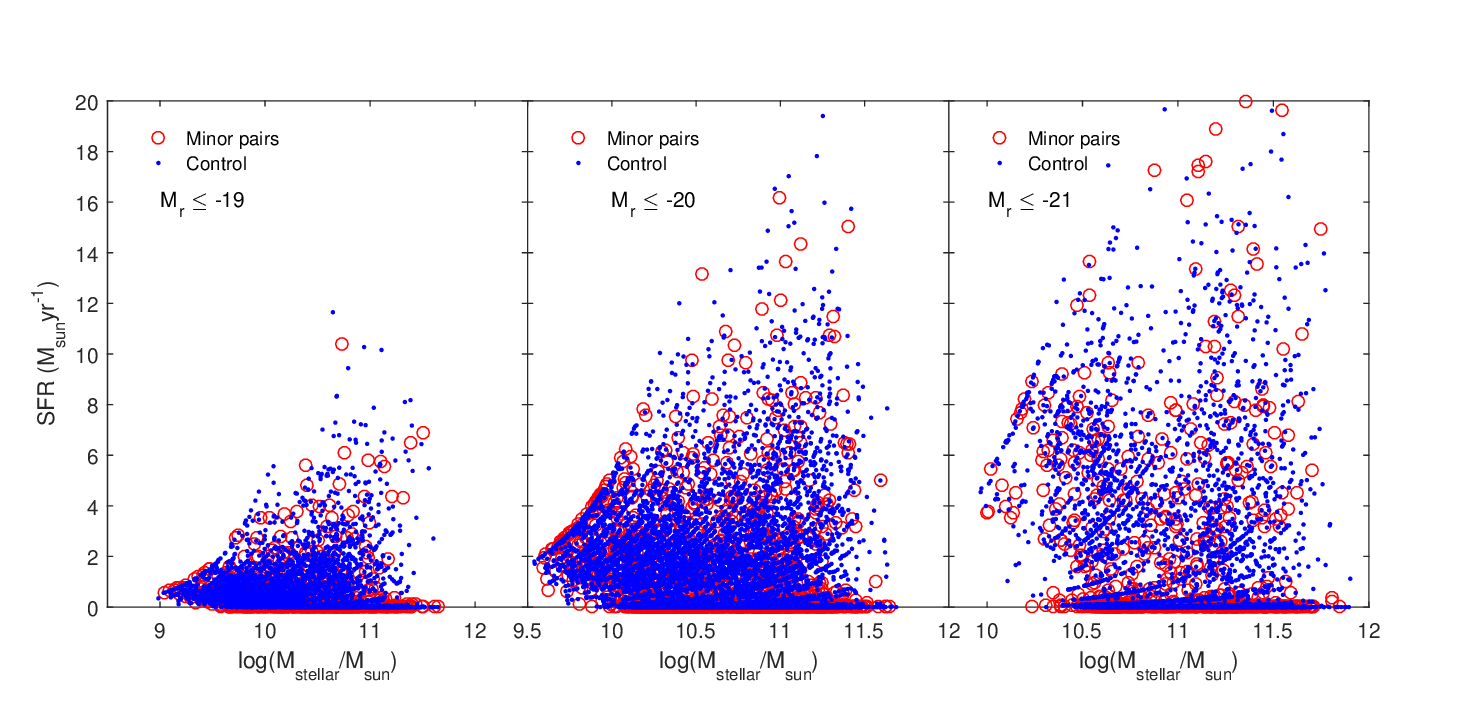}}}\
\caption{Three panels of this figure show the stellar mass - SFR
  relation of the minor pairs and their controls in the three volume
  limited samples.}
\label{masssfr}
\end{figure*}

\begin{figure*}

\resizebox{15cm}{6cm}{\rotatebox{0}{\includegraphics{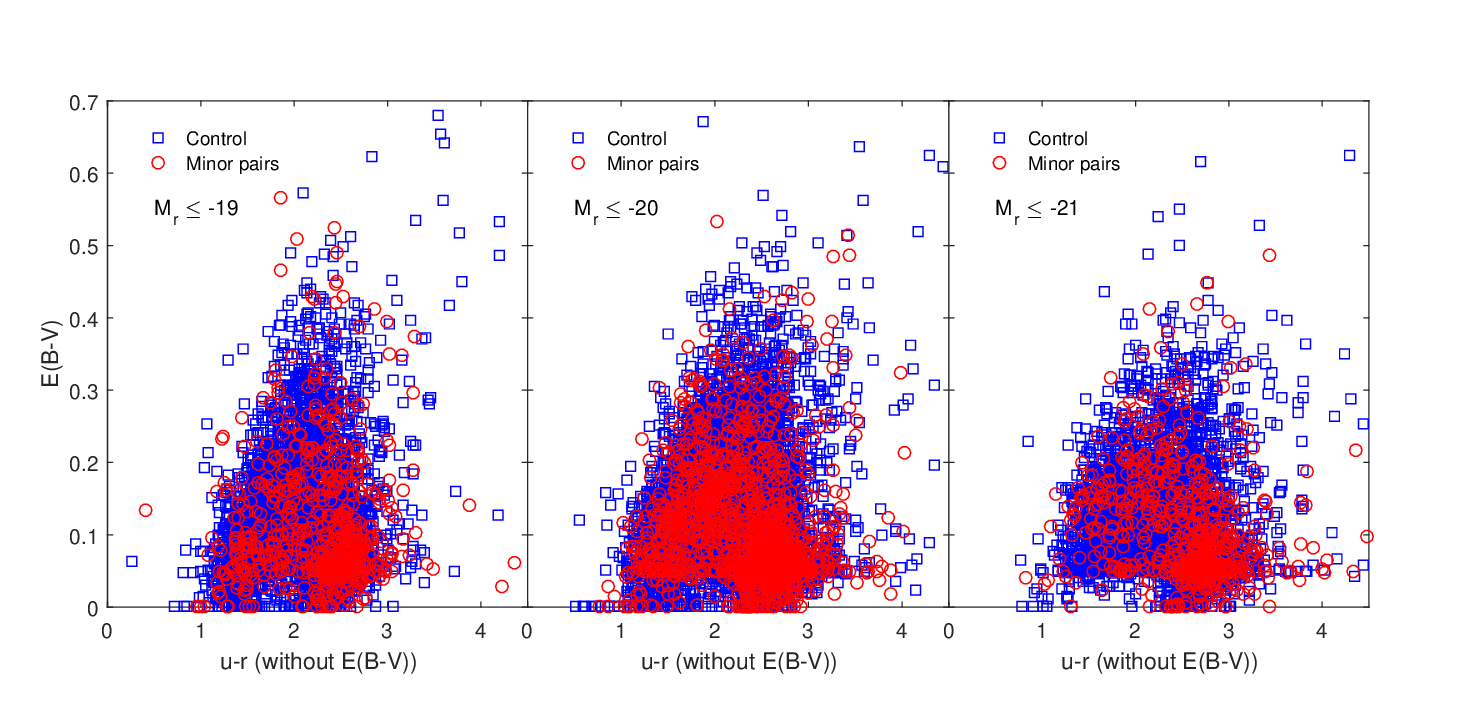}}}\
\caption{Different panels of this figure show the observed u-r colour
  versus E(B-V) colour excess in the three volume limited samples.}
\label{ebv}
\end{figure*}

\subsection{Possible limitations of the study}

We discuss about a few limitations of our study in this subsection.
It is challenging to obtain reliable photometry in blending systems
particularly for the low mass galaxies in paired systems. We use the
`clean' flag in the SDSS database to check whether the minor merger
systems in our samples have a reliable photometry. We show the numbers
of pairs with clean or unclean photometry as a function of projected
separation in the three volume limited samples in
\autoref{photo}. There are a small number of minor pairs present in
each sample for which both the members have unclean photometry. We see
that the relative abundance of such pairs do not depend on their
projected separations. Most of the galaxies in minor pairs have a
clean photometry in our samples.  However, the pairs with unclean
photometry may have some impact on our results. Considering this, we
also repeat our analysis with only the minor pairs with clean
photometry and find that our conclusions remain unchanged. The
quenching observed in minor pairs in our study is statistical and the
trends are identical in all three magnitude bins.

It is also challenging to measure the total flux of each galaxy in the
merger systems and calculate the individual SFRs accurately. We use
the SFRs derived from a catalogue which is based on the stellar
population synthesis (SPS) technique \citep{conroy09}. The SPS models
translate the observations (SED, magnitudes etc.) to physical
properties using a set of fit parameters in the model. We plot the
stellar mass versus SFR of the galaxies in minor pairs and their
control samples in all three volume limited samples in
\autoref{masssfr}. We find that the galaxies in these samples follow a
main sequence relation but with larger scatters at higher masses. This
may be related to the substantial uncertainties in the SPS
modelling. Such uncertainties may have some impact on our results.

The colour is also affected by the reddening or dust extinction in
galaxies. The additional reddening is often quantified with the E(B-V)
colour. It is determined by comparing the observed colors of galaxies
with the expected colors of an unreddened stellar population. This
colour excess affect the intrinsic properties like observed colour and
SFR of galaxies. A very large reddening for the galaxies in our sample
may in principle affect the results of our analysis. We show the
colour excess E(B-V) versus the observed (u-r) colour (without E(B-V)
correction) for the minor pairs and their control galaxies in all
three volume limited samples in \autoref{ebv}. We find that the colour
of the minor pairs and their controls are reasonable and similar in
all three samples. We do not expect these to affect our conclusions.


\begin{figure*}

\resizebox{15cm}{6cm}{\rotatebox{0}{\includegraphics{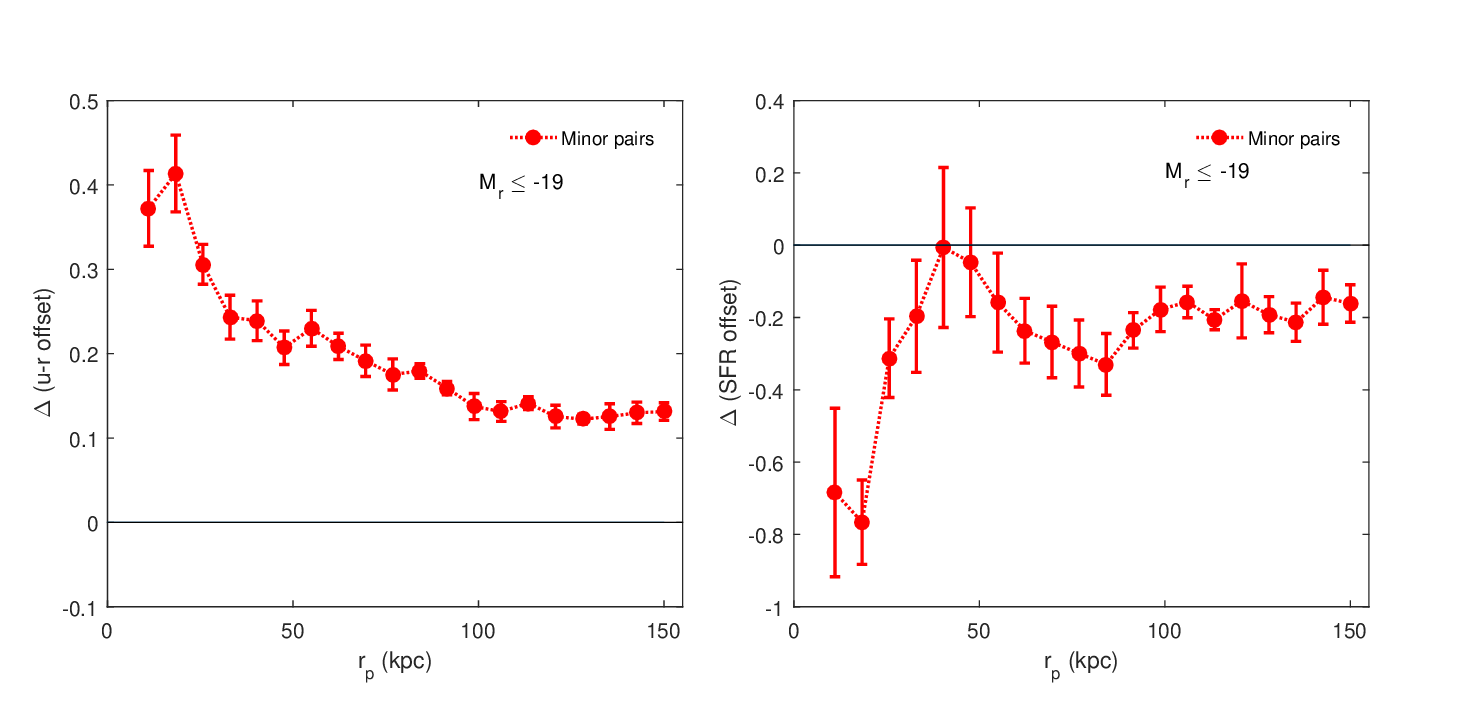}}}\
\caption{This plot shows color offset difference and SFR offset
  difference of minor pairs having clean photometry in $M_r \leq -19$
  magnitude bin. Here the SFR of minor pairs and their control matched
  sample is estimated from $H_{\alpha}$ emission line as discussed in
  \autoref{subsec:emm}.}
\label{off_final}
\end{figure*}

\subsection{Roles of photometric quality and SFR estimator}
\label{subsec:emm}

We address these limitations of our study by considering only the
galaxies with clean photometry and calculating the SFR of the minor
pairs using the $H_\alpha$ line. It is crucial to identify the
galaxies having reliable photometry for any analysis with galaxy
pairs. Some of the galaxies in our minor pair sample do not have
reliable photometry.  We only consider the galaxies with a reliable
photometry by using the 'clean' flag in the SDSS database. The
brightest magnitude bin in our sample may include some BCGs with their
satellites as minor pairs. It shows a higher degree of quenching
compared to the other two magnitude bins (\autoref{fig6}). The
inclusion of BCGs may be responsible for a higher quenching signal in
the brightest magnitude bin. Keeping this in mind, we only repeat our
analysis in the faintest magnitude bin ($M_r \leq -19$) considered in
our analysis. We now have $207$ minor pairs in this bin that have
clean photometry and $H_{\alpha}$ line information. The control
samples for these minor pairs are obtained following the same method
described in \autoref{sec:csample}.

Our results could be sensitive to different SFR estimators. We
calculate the SFR of minor pairs and their control matched sample
using the $H_{\alpha}$ emission line of these galaxies. The SFR of
these galaxies are estimated using the formula \citep{hopkins03},

\begin{equation}
SFR(M_{sun}yr^{-1})=4\pi
D_l^2S_{H_{\alpha}}\frac{10^{-0.4(r_{petro}-r_{fiber})}}{1.27 \times
  10^{34}}(\frac{S_{H_{\alpha}}}{2.86 \times S_{H_{\beta}}})^{2.114}
\end{equation}

Here $S_{H_{\alpha}}$, $S_{H_{\beta}}$ are stellar absorption
corrected $H_{\alpha}$ and $H_{\beta}$ fluxes respectively. The
$r_{petro}$, $r_{fiber}$ are the Petrosian and fiber magnitudes in the
$r$ band of SDSS and $D_l$ represents the luminosity distance of the
galaxies. The information about the spectral lines of minor pairs and
their control matched sample is obtained from the SDSS database by
matching their $SpecObjId$s. The cumulative median of the colour
offset difference and SFR offset difference are shown as a function of
projected separation in \autoref{off_final}. The results show a
positive value of colour offset difference and negative value of SFR
offset difference for nearly the entire pair separation range. We find
that our results do not change after we discard the galaxies with
unreliable photometry and employ a different SFR estimator for our
analysis.


\section{Conclusions}

We study the effects of tidal interactions on the SFR, and the dust
corrected $(u-r)$ colour of the minor pairs using a set of volume
limited samples from the SDSS. We first prepare our control samples by
matching the stellar mass and redshift and then compare the SFR and
$(u-r)$ colour in the minor pairs against their control samples of
isolated galaxies. The analysis shows that the SFR and $(u-r)$ colour
distributions of minor pairs significantly differ from their control
galaxies in all three volume limited samples. The null hypothesis can
be rejected at $>99\%$ confidence level. Both the $\Delta$$(u-r$
offset) and the $\Delta$($SFR$ offset) as a function of pair
separation indicate a quenching in minor pairs. The degree of
quenching decreases with the increasing pair separation upto $\sim 50$
kpc. The minor pairs in the brighter samples exhibit a higher
quenching at a fixed pair separation. The more luminous galaxies
preferentially reside in the denser environments where the galaxies
are known to be redder. This luminosity dependence indicates some role
of the environment in quenching the galaxies in minor pairs.

The control samples of the minor pairs must also be matched in the
environment besides the stellar mass and redshift. We repeat our
analysis with carefully designed control samples that are
simultaneously matched in stellar mass, redshift and the local
density. Interestingly, the SFR and $(u-r)$ colour distributions of
the minor pairs significantly differ from their control galaxies. The
null hypothesis can still be rejected at $>99\%$ confidence level. We
find that the $\Delta$$(u-r$ offset) and the $\Delta$($SFR$ offset) as
a function of pair separation still indicate a quenching in minor
pairs even after controlling their environment. The degree of
quenching is sensitive to the pair separation upto $\sim 50$ kpc. Our
results suggest that the minor interactions suppress the SFR and
enhance the $(u-r)$ colour in galaxies. The suppression of the SFR
decreases with the increasing pair separation but a non-zero
suppression is observed throughout the length scale probed. There is
observational evidence that SFR in paired galaxies may be influenced
by tidal interactions up to $150$ kpc \citep{scudder12b, patton13}.

 The present analysis indicates that the minor interactions suppress
 the SFR and enhance the $(u-r)$ colour in galaxies. This result
 seemingly contradicts some of the findings reported in
 \citet{scudder12b}. This difference most likely arises due to the
 different treatment of environments. We use the density estimates in
 three-dimension using redshift information whereas \citet{scudder12b}
 use the projected densities for their work. Our pair sample is
 defined within a volume limited sample whereas the pair sample in
 \citet{scudder12b} was defined within a flux limited sample. Most of
 the selection biases are taken into account in a volume limited
 sample. The local density can be estimated more reliably within a
 volume limited sample. We would like to mention here that we indeed
 observe an enhanced star formation in interacting major pairs
 \citep{das21b, das23} in our analysis with volume limited
 samples. However, we observe an opposite trend for the minor pairs.

We also analyze the morphology of the minor pairs by cross-matching
the paired galaxies with the Galaxy Zoo and Galaxy Zoo2. The
Spiral-Spiral and Spiral-Elliptical combinations are more frequently
observed than the Elliptical-Elliptical combinations in the minor
pairs. We note that only $\sim 1\%$ galaxies have a dominant bulge,
$4\%-7\%$ galaxies host a bar, and $5\%-10\%$ galaxies show the AGN
activity in the minor pairs. It suggests that the bar, bulge and AGN
activity do not take a leading role in quenching the galaxies in the
minor pairs.

The minor interactions are believed to trigger a mild enhancement in
the star formation activity in the past. However, we do not find such
enhancement in the star formation activity in the minor pairs in the
local Universe. Our analysis indicates that the minor interactions in
the present Universe initiate quenching in galaxies. The degree of
quenching decreases with the increasing pair separation.  A
significant number of galaxies in the minor pairs have a stellar mass
above $3 \times 10^{10}\,M_{\odot}$ which are intrinsically redder and
less star-forming \citep{kauffmann03a}. We propose that the more
massive members in the minor pairs may curtail their star formation
through mass quenching \citep{binney04, birnboim03, dekel06, das21} or
major merger-driven quenching \citep{gabor, montero}. They possibly
quench the star formation in their less massive companions at a later
stage by stripping away the gas and leading them to starvation
\citep{larson80,somerville99,kawata08}. It should be noted that the
control galaxies in our analysis are matched in stellar mass. So any
mass driven quenching would be equally effective in the control
matched samples of isolated galaxies. However any quenching induced by
the interactions would be present only in the minor paired galaxies.

We carefully select the control galaxies by closely matching the
redshift, stellar mass and environment. It eliminates the biases in
our results. Any other systematic biases should equally affect both
the samples and should not be a matter of concern here. However, a few
caveats remain in our analysis. All the galaxies in close pairs may
not be undergoing interactions. Some pairs may be approaching each
other and are yet to experience an encounter. Also some of the
selected pairs may not be close in three dimensions due to the chance
superposition in the high-density regions like groups and clusters
\citep{alonso04}. Nevertheless, these caveats plague all previous
analysis of galaxy pairs with observational data.

  Any stellar mass incompleteness present in the pair
  and control samples would spoil the comparisons between their colour
  and SFR. We also apply stellar mass limit to our volume limited
  samples and repeat our entire analysis. We find that our conclusions
  remain unchanged in such an analysis.

We note that some galaxies in our minor pair sample have unreliable
photometry. The SFR estimates used in our analysis are derived from
the SED fitting. These estimates have large uncertainities at higher
stellar masses. We also notice a larger quenching signal for the
brightest magnitude bin in our analysis (\autoref{fig6}). This
indicates that the brightest magnitude bin may contain a significant
number of BCGs with their quenched satellite galaxies. Keeping these
issues in mind, we select only the galaxies with clean photometry in
the faintest magnitude bin. We estimate the SFR of these minor pairs
and their control matched sample using the $H_{\alpha}$ line in these
galaxies. We repeat our analysis to calculate the SFR and colour
offsets in the faintest magnitude bin. The results show that we still
obtain a quenching signal in the faintest magnitude bin. This suggests
that the quenching phenomena in minor pairs in our work can not be
explained by the unclean photometry, contaminations from BCGs and
differences between the SFR estimators.

Observations indicate that the star formation in galaxies peaked at a
redshift of $z\sim2-3$ \citep{tran10, forster20, gupta20}. This epoch
is often referred as the ``cosmic noon''. The minor interactions may
trigger star formation in galaxies in the early stages of their
evolution. The tidal interactions are more effective in inducing star
formation during this period due to the lack of stability in the
galaxies \citep{tissera02}. The minor interactions may have
contributed significantly to the rapid rise in cosmic star formation
rate during the ``cosmic noon''. The cosmic star formation rate
declines sharply between $z=1$ to $z=0$ \citep{madau96}.

We find that a significant number of minor pairs in our sample have a
very low SFR. These quiescent galaxies could be quenched due to
multiple reasons. For example, the environment may have some role in
such quenching. The presence of bar, dominant bulge and AGN may also
play some roles in quenching the galaxies in minor pairs. However,
they do not explain the quenching in most minor pairs. We expect the
more massive companion to be mass quenched or merger quenched in most
minor pairs. We propose that the less massive companion may experience
a satellite quenching at a later stage of evolution \citep{vanden,
  wright22}. However, these alternate scenarios can not be verified in
this work and require further studies. The quenching by minor
interactions in the present Universe would contribute significantly to
the build-up of the red sequence and the observed bimodality. We
propose that this should be taken into account while modelling the
observed bimodality.

\section{ACKNOWLEDGEMENT}
We sincerely thank an anonymous reviewer for the insightful comments
and suggestions that helped us to improve the draft. The authors thank
the SDSS team for making the data publicly available. BP would like to
acknowledge financial support from the SERB, DST, Government of India
through the project CRG/2019/001110. BP would also like to acknowledge
IUCAA, Pune for providing support through associateship programme. SS
acknowledges IISER Tirupati for support through a postdoctoral
fellowship.

Funding for the SDSS and SDSS-II has been provided by the Alfred
P. Sloan Foundation, the Participating Institutions, the National
Science Foundation, the U.S. Department of Energy, the National
Aeronautics and Space Administration, the Japanese Monbukagakusho, the
Max Planck Society, and the Higher Education Funding Council for
England. The SDSS website is http://www.sdss.org/.

The SDSS is managed by the Astrophysical Research Consortium for the
Participating Institutions. The Participating Institutions are the
American Museum of Natural History, Astrophysical Institute Potsdam,
University of Basel, University of Cambridge, Case Western Reserve
University, University of Chicago, Drexel University, Fermilab, the
Institute for Advanced Study, the Japan Participation Group, Johns
Hopkins University, the Joint Institute for Nuclear Astrophysics, the
Kavli Institute for Particle Astrophysics and Cosmology, the Korean
Scientist Group, the Chinese Academy of Sciences (LAMOST), Los Alamos
National Laboratory, the Max-Planck-Institute for Astronomy (MPIA),
the Max-Planck-Institute for Astrophysics (MPA), New Mexico State
University, Ohio State University, University of Pittsburgh,
University of Portsmouth, Princeton University, the United States
Naval Observatory, and the University of Washington.

\section{DATA AVAILABILITY}
The data underlying this article are publicly available at
https://skyserver.sdss.org/casjobs/.

\label{lastpage}
\end{document}